\documentclass[fleqn,usenatbib]{mnras}

\usepackage{newtxtext,newtxmath}

\usepackage[T1]{fontenc}

\newcommand{\bl}[1]{\mbox{\boldmath$ #1 $}}

\DeclareRobustCommand{\VAN}[3]{#2}
\let\VANthebibliography\thebibliography
\def\thebibliography{\DeclareRobustCommand{\VAN}[3]{##3}\VANthebibliography}

\usepackage{graphicx}	
\usepackage{amsmath}	
\graphicspath{{Figures/}}

\title[Gravito-Viscous Evolution of Young Disks]{Testing Models of Disk Formation and Gravito-Viscous Evolution Against Observations of Young Protoplanetary Disks}

\author[F. A. Gerosa et al.]{
Fabiola Antonietta Gerosa,$^{1}$\thanks{E-mail: f.gerosa@ucl.ac.uk}
Paola Pinilla,$^{1}$
Eduard I. Vorobyov $^{2,3}$
and James McKevitt$^{1,3}$
\\
$^{1}$Mullard Space Science Laboratory, University College London, Holmbury St Mary, Dorking, Surrey, RH5 6NT, UK\\
$^{2}$ Institut für Astro- und Teilchenphysik, Universität Innsbruck, Technikerstraße 25, 6020 Innsbruck, Austria\\
$^{3}$University of Vienna, Department of Astrophysics, Türkenschanzstrasse 17, 1180, Vienna, Austria
}

\date{Accepted XXX. Received YYY; in original form ZZZ}

\pubyear{\the\year{}}

\begin{document}
\label{firstpage}
\pagerange{\pageref{firstpage}--\pageref{lastpage}}
\maketitle

\begin{abstract}
Young protoplanetary disks have received increasing attention in recent years, due to observations revealing the presence of substructures in disks younger than one million years, suggesting that  planet formation may begin earlier than previously thought. Other recent observations have provided new insights into the evolution of gas and dust in these early stages, enabling comparisons with
evolved Class II disks. Yet the physical processes governing the earliest stages of disk evolution remain poorly understood.
In this paper, we present two-dimensional thin-disk hydrodynamical simulations of the formation and early evolution of protoplanetary disks, under the influence of self-gravity and turbulent viscosity alone. We explore various properties of disks forming from the collapse of clouds with a range of initial masses and we compare our results with surveys of protoplanetary disks across multiple evolutionary stages. Our simulations broadly reproduce key observational constraints, including the mm fluxes of Class 0 disks and the dust disk radii of later stages.
However, our models do not explain the observed gas and dust disk masses of Class I and II disks: while they account for the most massive observed disks, they fail to reproduce the least massive ones.  We discuss that simulations incorporating the influence of the star-forming environment might be needed to resolve this discrepancy.

\end{abstract}




\section{Introduction}
Stars and protoplanetary disks form from the gravitational collapse of a molecular cloud. Protoplanetary disks are mostly composed of gas, but also contain a small portion of dust. This material can grow to form planets and other smaller bodies. Throughout their lifetime, young stellar objects (YSOs) undergo different evolutionary phases. In the initial phase, known as Class 0, the disk is still embedded in the molecular cloud and jets are usually present due to ongoing accretion from the envelope. This phase is 
quickly followed by the Class I stage, in which the disk is surrounded by a less dense envelope. 
The Class II stage begins when the envelope has finally disappeared and can last for several million years. The disk eventually dissipates to leave a planetary system \citep{andre2002}.

The emergence of planetary systems from protoplanetary disks around young stars has now been observed directly in two systems, PDS 70~\citep{keppler2018,muller2018,Haffert2019} and WISPIT2~\citep{vanCapelleveen2025,Lawlor2026}. Exoplanetary systems (in which the disk has already dispersed) have been extensively observed in the last 30 years. As of today, more than 6200 exoplanets have been discovered, showing extreme diversity and demonstrating that  planet formation in protoplanetary disks must be a common process. However, many aspects of this process are still unknown, including when planet formation may occur throughout the lifetime of a protoplanetary disk. The aforementioned PDS 70 and WISPIT2 systems are rather old (5-7~Myr), so they do not provide clear insights on how early their planets started to form. 

The presence of substructures in protoplanetary disks can provide insight into the possibility of early planet formation. Substructures are quite ubiquitous in Class II disks~\citep{Andrews2018,Teague2025}. 
Their appearance could be caused by various mechanisms (e.g. ice lines, instabilities, photoevaporation, magneto-hydrodynamic processes; see \citet{Bae2023} for a review), including the presence in these protoplanetary disks of already formed planetary companions.
Therefore, the detection of substructures in Class II disks could indicate the presence of giant planets in these systems. However, substructures seem to be less common in Class 0 and I disks. The eDisk project inspected 19 Class 0 and I disks and found very few substructures in them \citep{Ohashi2023}. They concluded that most of the planet formation process should occur rapidly in between the late Class I and early Class II stages. However, recent surveys have shown that substructures are already common in Class I disks~\citep{Shoshi2025,Hsieh2025,Bhowmik2026}, suggesting that the planet formation process may begin earlier, at the start of the Class I stage or even before.  Moreover, ~\citet{Nazari2025} found that massive planets ($\lesssim 10^2 M_{\oplus}$), even if already formed in Class 0 and Class I disks, would not be opening gaps in these early and warm disk phases. Therefore, even if they were present, young planets would be difficult to detect in the embedded stages. The suggestion of early planet formation is also supported by the  "mass budget problem", whereby the dust content in Class II disks is insufficient for planet formation~\citep{Manara2018}. Disks may be more massive than estimated due to their optical thickness. However, planet formation in earlier, more massive disk phases is still a possible solution to this problem, at least for forming the most massive planets. 

Observations of disks in the early stages are challenging, in particular because of the increased optical depth and extinction due to the envelope surrounding them. Nonetheless, several efforts have been made to constrain the disk properties. They include observing at longer wavelengths to mitigate high optical depth effects~\citep{Tobin2016}, combining high- and low-angular resolution observations to distinguish the envelope from the disk emission \citep{Ohashi2023}, and detecting different  combinations of molecular lines that allow tracing the gas disk masses \citep{zhang_alma_2025}. These studies have allowed us to have the first observational evidence that a significant gas evolution has happened between the Class 0/I stages and the early Class II objects, slowing significantly during the later Class II phase \citep{zhang_alma_2025}. Dust evolution might also start in early stages, suggested by the presence of substructures in various Class I disks, as discussed above, and by the small spectral index values found in the envelopes and disks of young stellar systems \citep{Miotello2014,Cacciapuoti2023}. 

For this reason, it is essential to better constrain how disks form and evolve from the collapse of molecular clouds and to compare theoretical predictions with available observations. This would help to identify the key properties and physical processes that operate in the earliest evolutionary stages and shape the Class II phase, influencing the possible outcome of the planet formation process. 
Different previous studies have addressed the early phases of disk formation in cloud collapse simulations, focusing on different mechanisms driving protostellar accretion and disk evolution. Some examples include previous thin-disk or 3D works on the masses and sizes of embedded disks~\citep[e.g.,][]{Vorobyov2011,2014Machida,Hennebelle2016,2020Hennebelle,Vorobyov2024,2025ApJ...990...95K}, population synthesis of disks formed in hydrodynamical or magnetohydrodynamical (MHD) simulations of star forming clusters~\citep[e.g.,][]{Kuffmeier2017,Bate2018,Lebreuilly2021,Lebreuilly2024}, studies considering semi-analytic models of envelope infall into the disk~\citep[e.g.,][]{Hueso2005,Visser2009,Kimura2016K}, and others focusing on dust dynamics and growth during these early stages~\citep[e.g.,][]{2020Lebreuilly, 2020Tsukamoto, Appelgren2023,Vorobyov2025,Bate2026} 
However, only a few studies have numerically modeled gravito-viscous disks across different evolutionary stages. In particular, multi-dimensional simulations that couple gas and dust with self-consistent growth and drift from the embedded phases to Class II, and their statistical comparison with recent surveys, are still missing.

In this paper, we investigate the formation and evolution of protoplanetary disks. We perform two-dimensional thin-disk numerical hydrodynamics simulations of systems composed of an infalling molecular cloud, forming a star and a circumstellar disk evolving under the influence of disk self-gravity and viscosity. We have selected different masses of molecular clouds for our simulations and we have conducted a partial exploration of other initial conditions and parameters. Various properties of these disks are then computed and compared to those of observed disks at different stages of evolution.

This paper is organised as follows. In Section \ref{sec:methods}, we introduce the numerical method used for this work, the selected simulation setup and the analysis approach adopted. In Section \ref{sec:Results}, we present the results  from our simulations and the comparison with observations of disks from various surveys. Section \ref{sec:discussion} discusses various mechanisms that may be important for the evolution of protoplanetary disks, but which are not considered in this work. Finally, in Section \ref{sec:conclusion} we conclude the paper summarizing the results of this study.

\section{Methods}
\label{sec:methods}
\subsection{FEOSAD}
\label{sec:FEOSAD} 

Numerical simulations of disk formation and evolution were carried out using the Formation and Evolution Of Stars And Disks (FEOSAD) numerical hydrodynamics code. FEOSAD computes the coevolution of gas and dust, beginning from the cloud collapse stage and proceeding through the main star and disk evolution stages. The equations of hydrodynamics were solved in the thin-disk limit, the justification of which is provided in \citet{VorobyovBasu2009}. The detailed description of the code is presented in \citet{2018VorobyovAkimkin}, with modifications to include the back-reaction of dust on gas and the Stokes drag regime in \citet{Vorobyov2023}. Below, we provide a concise description of the main constituents of the code.

\subsubsection{Gas dynamics}
The gas dynamics is followed by numerically solving the continuity equation for the gas surface density, and the equations for the momentum and internal energy per unit surface area. The forces that determine the disk's dynamics include gravity (both central source and disk self-gravity), viscosity, and friction between gas and dust. 
The disk's internal energy is in turn determined by viscous heating, radiative heating (including the radiation of a nascent star and background radiation), dust radiative cooling, and adiabatic work. The corresponding equations in the thin-disk limit are as follows.
\begin{equation}
\label{eq:cont}
\frac{{\partial \Sigma_{\rm g} }}{{\partial t}}   + \nabla  \cdot 
\left( \Sigma_{\rm g} {\bl v} \right) = 0,  
\end{equation}
\begin{eqnarray}
\label{eq:mom}
\frac{\partial}{\partial t} \left( \Sigma_{\rm g} {\bl v} \right) +  \nabla \cdot \left( \Sigma_{\rm
g} {\bl v} \otimes {\bl v} \right)  =   - \nabla {\cal P}  &+& \Sigma_{\rm g} \, {\bl g} + \nonumber
\\ 
&+& \nabla \cdot \mathbf{\Pi}  - \Sigma_{\rm d,gr} {\bl f},
\end{eqnarray}
\begin{equation}
\frac{\partial e}{\partial t} +\nabla \cdot \left( e {\bl v} \right) = -{\cal P} 
(\nabla \cdot {\bl v}) -\Lambda +\Gamma + 
\left(\nabla {\bl v}\right):\Pi, 
\label{eq:energ}
\end{equation}
where  $\Sigma_{g}$ and $e$ are the gas surface density and the internal energy per surface area, respectively, ${\bl v}=v_r\hat{{\bl r}}+v_\phi \hat{{\bl \phi}}$ is the gas velocity in the disk midplane, $\cal{P}$ is the pressure, integrated in the vertical direction using the ideal equation of state ${\cal P}=(\gamma-1) e$ with $\gamma=7/5$, and  ${\bl f}$  is the drag force per unit mass between gas and dust. The gravitational acceleration in the disk plane ${\bl g}$ takes into account the gravity of the central star, and the gas and dust self-gravity in the disk, which is found by solving the integral form for the potential using the convolution method as laid out in \citet{BT1987} and \citet{Vorobyov2024}. Stellar motion in response to the gravity of the surrounding medium, often modeled in the non-inertial frame via an indirect potential~\citep{Crida2025,Crida2025r}, is not considered in this study,  as the presence of an extended envelope may lead to instability and, in extreme cases, to code termination~\citep{Regaly2017}.
The radiative cooling  and  heating rates are denoted by $\Lambda$ and $\Gamma$ as described in \citep{DongVorobyov2016}.
The Rosseland and Planck mean opacities were taken from \citet{Semenov2003}.

Turbulent viscosity is parameterized using the  \citet{1973ShakuraSunyaev} approach as $\nu = \alpha c_{\mathrm{s}} H_{\mathrm{g}}$, where $\nu$ is the kinematic viscosity that enters the viscous stress tensor $\Pi$, $c_{\mathrm{s}}$ is the sound speed, and $H_{\mathrm{g}}$ is the gas vertical scale height, calculated using an assumption of local hydrostatic balance. 
We assume that turbulence is  isotropic, and $\alpha$ represents not only the efficiency of mass and angular transport in the disk plane due to turbulence, but also the efficiency of dust growth and settling in the monodisperse dust growth model described in detail in \citet{2018VorobyovAkimkin} and summarized in Subsection \ref{subsubs:mono}.  

\subsubsection{Dust dynamics}

FEOSAD computes the dynamics of two dust populations: (i) small dust, which are grains with a size\footnote{Here and further in the text by size of dust grains we mean their radius.} between $a_{\rm min}=5\times 10^{-3} \ \mu \rm m$  and $a_{*} = 1 \ \mu \rm m$ and (ii) grown dust ranging in size from $a_{*}$ to a maximum $a_{\rm max}$. The latter is variable in space and time, depending on the efficiency of dust growth, drift, and fragmentation.  Dust in both populations is distributed over size according to a simple power law: $N(a) = C \cdot a^{ - {\rm p}}$, where $N(a)$ is the number of dust particles per unit dust size, $C$ is a normalization constant, and ${\rm p} = 3.5$ is the power index, which is kept constant during the considered disk evolution period. A more sophisticated approach requires solving for the Smoluchowski equation for multiple dust bins and is beyond the scope of the present study.

We solve the continuity equation separately for the grown and small dust populations. However, the momentum equation is solved only for the grown dust, because small dust is assumed to be dynamically coupled to the gas.  The system of hydrodynamics equations for dust in the zero-pressure limit is written as follows,
\begin{equation}
\label{eq:contDsmall}
\frac{{\partial \Sigma_{\rm d,sm} }}{{\partial t}}  + \nabla \cdot 
\left( \Sigma_{\rm d,sm} {\bl v} \right) = \nabla  \cdot \left( D \Sigma_{\rm g} \nabla\left(\frac{\Sigma_{\rm d,sm}}{\Sigma_{\rm g}} \right)\right) - S(a_{\rm max}), 
\end{equation}
\begin{equation}
\label{eq:contDlarge}
\frac{{\partial \Sigma_{\rm d,gr} }}
{{\partial t}}  + \nabla  \cdot 
\left( \Sigma_{\rm d,gr} {\bl u} \right) = 
 \nabla  \cdot \left( D \Sigma_{\rm g} \nabla\left(\frac{\Sigma_{\rm d,gr}}{\Sigma_{\rm g}} \right)\right) +  S(a_{\rm max}),
\end{equation}
\begin{equation}
\label{eq:momDlarge}
\frac{\partial}{\partial t} \left( \Sigma_{\rm d,gr} {\bl u} \right) +  \nabla \cdot \left( \Sigma_{\rm d,gr} {\bl u} \otimes {\bl u} \right)  =   \Sigma_{\rm d,gr} \, {\bl g} + \Sigma_{\rm d,gr} \, {\bl f} 
 + S(a_{\rm max}) {\bl v},
\end{equation}
where $\Sigma_{\rm d,sm}$ and $\Sigma_{\rm d,gr}$ are the surface densities of small and grown dust, respectively, and ${\bl u}$ are the planar components of the grown dust velocity.  
The first term on the right hand side of  Eqs.~\eqref{eq:contDsmall} and~\eqref{eq:contDlarge} describes dust turbulent diffusion. The coefficient of turbulent diffusion $D$ is related to the kinematic viscosity $\nu$ as $D=\nu/(1+\mathrm{St}^2)$, where $\mathrm{St}$ is the Stokes number, which is the ratio between the orbital timescale and the stopping time of the particle. 

The drag force (per unit mass) links grown dust with gas and can be written as \citep{Weidenschilling1977}:
\begin{equation}
    {{\bl f}} = \dfrac{1} {2 m_{\rm d}} C_{\rm D} \, \sigma \rho_{\rm g} ({{\bl v_p}} - {{\bl u_p}}) |{{\bl v_p}} - {{\bl u_p}}|,
\label{eq:friction}
\end{equation}
where $\sigma$ is the dust grain cross section, $\rho_{\rm g}$ the volume density of gas, $m_{\rm d}$ the mass of a dust grain, and $C_{\rm D}$ the dimensionless friction parameter. The latter is described in details in \citet{Vorobyov2023} and is based on the works of \citet{Henderson1976} and \citet{Stoyanovskaya2020}.
The use of the Henderson friction coefficient allows us to treat the drag force in the Epstein and Stokes regimes, depending on the local conditions and dust size. We are interested in the dynamics of dust grains that are the main mass carriers. Therefore, we use the maximum size of dust grains $a_{\rm max}$ when calculating the values of $\sigma$ and $m_{\rm d}$ in Equation~(\ref{eq:friction}). This choice may slightly underestimate the friction force ${\bl f}$.

\subsubsection{Monodisperse dust growth scheme}
\label{subsubs:mono}
As the disk forms and evolves, dust transits from small to grown dust population. This process is described by the term $S(a_{\rm max})$, which enters the equations for the dust component and represents the conversion rate between small and grown dust populations. 
For a mathematical and graphical description of small-to-grown dust conversion process we refer to~\citet{Molyarova2021}. Here, we note that $S(a_{\rm max})$ depends only on the local maximal size of dust grains $a_{\rm max}$, which in turn is determined by the rate of collisional growth and fragmentation, and also by the flow of dust through the disk. The equation describing the dynamical evolution of $a_{\rm max}$ is therefore as follows
\begin{equation}
{\partial a_{\rm max} \over \partial t} + ({\bl u} \cdot \nabla ) a_{\rm max} = \cal{D},
\label{eq:dustA}
\end{equation}
where the rate of dust growth due to collisions and coagulation is calculated in the monodisperse approximation \citep{2012Birnstiel}
\begin{equation}
\cal{D} = {\rho_{\rm d} \mathit{u}_{\rm rel} \over \rho_{\rm s}}.
\end{equation}
This rate depends on the total volume density of dust $\rho_{\rm d}$, the dust material  density $\rho_{\rm s} = 2.24$~g~cm$^{-3}$ \citep{Draine2001}, and the relative velocity of particle-to-particle collisions defined as $\mathit{u}_{\rm rel} = (\mathit{u}_{\rm th}^2 + \mathit{u}_{\rm turb}^2)^{1/2}$, where $\mathit{u}_{\rm th}$ and $\mathit{u}_{\rm turb}$ account for the Brownian and turbulence-induced local motion, respectively.  The dust-to-dust collision velocity owing to turbulence is computed following the model of turbulent eddies proposed in \citet{Ormel2007} 
\begin{equation}
    u_{\rm{turb}} = \sqrt{{\frac{3 \alpha}{\mathrm{St}+\mathrm{St}^{-1}}}} c_{\rm s}.
    \label{turb_vel}
\end{equation}
When calculating $\rho_{\rm d}$, we take into account dust settling by calculating the effective scale height of grown dust  $H_{\rm d}$ as
\begin{equation}
    H_{\rm d} = H_{\rm g} \sqrt{ {\alpha \over \alpha+\mathrm{St} } }.
    \label{eq:dust-scale-height}
\end{equation}

Dust growth in FEOSAD is limited by  drift and collisional fragmentation barriers. The former is accounted for self-consistently when computing the dynamics of grown dust, while the latter is taken into account by calculating the characteristic fragmentation size in the Epstein regime as \citet{2016Birnstiel}: 
\begin{equation}
    a_{\rm frag}=\frac{2\Sigma_{\rm g} \mathit{u}_{\rm frag}^2}{3\pi\rho_{\rm s} \alpha c_{\rm s}^2},
\label{eq:afrag}
\end{equation}
where $\mathit{u}_{\rm frag}$ is the fragmentation velocity, namely, a threshold value of the relative velocity of dust particles at which collisions result in fragmentation rather than coagulation. 
If $a_{\rm max}$ becomes greater than $a_{\rm frag}$, we stop the growth of dust and set $a_{\rm max} = a_{\rm frag}$.

\subsubsection{Initial and boundary conditions}
\label{Sect:initial}
Simulations start from the gravitational collapse of a flattened prestellar cloud core, consisting initially of gas and small dust. 
The initial dust-to-gas ratio is 1:100. 
The gas surface density and angular velocity of the natal prestellar core are distributed as follows \citep{1997Basu}:
\begin{equation}
    \Sigma_{\rm g}(r)=\frac{r_0\Sigma_{\rm 0}}{\sqrt{r^2+r_0^2}},
\label{eq:coredens}
\end{equation}
\begin{equation}
    \Omega_{\rm g}(r)=2\Omega_{\rm 0}\bigg(\frac{r_0}{r}\bigg)^2\left[\sqrt{1+\left(\frac{r}{r_0}\right)^2}-1\right],
\label{eq:corevel}
\end{equation}
where $\Sigma_{\rm 0}$ is the surface density  and $\Omega_{\rm 0}$ is the angular velocity, both defined at the center of the cloud core, and $r_0$ is the radius of the near-uniform region (plateau) in the center of the core. By default, every cloud has a constant ratio $r_{\rm out}/r_0=6$, where $r_{\rm out}$ is the core's outer radius. To initiate the collapse, the initial density distribution receives a positive perturbation with an amplitude of 10\%. 

As the core contracts gravitationally, it spins up and a centrifugal disk forms when the in-spiralling gas hits the centrifugal barrier. Subsequently, the disk grows in size and mass owing to infall from the cloud, while the central star gains mass via accretion through the inner computational boundary $r_{\rm sc}$, at which a "sink-cell" is introduced. Free mass exchange (inflow and outflow) across the sink-disk interface is introduced at the inner boundary condition (see~\citet{2018VorobyovAkimkin} for details). 
The outer boundary condition allows free mass outflow, but mass inflow from outside the computational domain is prohibited.

\subsection{Model setup}
\begin{table}
	\centering
	\caption{Summary of the simulations}
	\label{tab:table_simulations}
	\begin{tabular}{ccccc} 
		\hline
		ID & Cloud mass (M$_\odot$) & $\beta$ & $\alpha$ & $\varv_{\rm{frag}}$ (m~s$^{-1})$\\
		\hline
		M0.15 & 0.15 & 10$^{-3}$ & 10$^{-3}$ & 5\\
        M0.275 & 0.275 & 10$^{-3}$ & 10$^{-3}$ & 5\\
		  M0.40 & 0.40 & 10$^{-3}$ & 10$^{-3}$ & 5\\
		M0.65 & 0.65 & 10$^{-3}$ & 10$^{-3}$ & 5\\
            M0.90 & 0.90 & 10$^{-3}$ & 10$^{-3}$ & 5\\
            M1.15 & 1.15 & 10$^{-3}$ & 10$^{-3}$ & 5\\
            M1.40 & 1.40 & 10$^{-3}$ & 10$^{-3}$ & 5\\
            M1.65 & 1.65 & 10$^{-3}$ & 10$^{-3}$ & 5\\
            M1.90 & 1.90 & 10$^{-3}$ & 10$^{-3}$ & 5\\
            M2.15 & 2.15 & 10$^{-3}$ & 10$^{-3}$ & 5\\
            M2.40 & 2.40 & 10$^{-3}$ & 10$^{-3}$ & 5\\
            M2.65 & 2.65 & 10$^{-3}$ & 10$^{-3}$ & 5\\
            $\beta0.5\_$M0.40 & 0.40 & 5$\cdot$10$^{-4}$ & 10$^{-3}$ & 5\\
             $\beta0.5\_$M1.40 & 1.40 & 5$\cdot$10$^{-4}$ & 10$^{-3}$ & 5\\
             $\beta2\_$M0.40 & 0.40 & 2$\cdot$10$^{-3}$ & 10$^{-3}$ & 5\\
             $\beta2\_$M1.40 & 1.40 & 2$\cdot$10$^{-3}$ & 10$^{-3}$ & 5\\
             $\alpha$-2$\_$M0.40 & 0.40 & 10$^{-3}$ & 10$^{-2}$ & 5\\
             $\alpha$-2$\_$M1.40 & 1.40 & 10$^{-3}$ & 10$^{-2}$ & 5\\
             $\varv_{\rm{frag}}10\_$M0.40 & 0.40 & 10$^{-3}$ & 10$^{-3}$ & 10\\
             $\varv_{\rm{frag}}10\_$M1.40 & 1.40 & 10$^{-3}$ & 10$^{-3}$ & 10\\
		\hline
	\end{tabular}
\end{table}
We perform 12 simulations, varying the initial mass of the infalling cloud in the interval [0.15 - 2.65]~M$_{\odot}$. The different cloud masses result from selecting increasing values of the cloud outer radius, i.e. varying $r_{\rm out}$.
All the other initial parameters are kept fixed in the fiducial simulations. In particular, we set the $\alpha$ parameter to $10^{-3}$. This value represents an intermediate regime between strong MRI-driven turbulence and laminar flow dominated by magnetic disk winds \citep{Bai2013}. It is also one of the typical levels of turbulence measured from observations using different techniques \citep{2023Rosotti}. The ratio of rotational-to-gravitational energy is set to $\beta = 10^{-3}$. This value is within the limits inferred from observations of N$_2$H$^+$ in prestellar cloud cores \citep{caselli_dense_2002}, but is towards the lower end of the range. This choice is necessary for simulations involving massive prestellar cores, which form massive and extended disks that are prone to fragmentation unless the core's angular momentum is low. Studying disk fragmentation is outside the scope of this paper. 
Finally, we set the fragmentation velocity $\varv_{\rm{frag}}$ to 5~m~s$^{-1}$, an intermediate value between the silicate dust and icy dust measured thresholds~\citep{poppe1997,wurm1998,poppe2000,gundlach2015}. The dependence of the results on these three parameters is  explored in Subsection \ref{subsec:Parameter_exploration}, where 8 extra simulations with different values of $\alpha$, $\beta$ and $\varv_{\rm{frag}}$ are presented. The parameters used for the 20 simulations presented in this paper are summarized in Table~\ref{tab:table_simulations}.

The grid resolution is 256$^2$, such that the disk scale height at half of the gas disk radius is covered by about two cells.  
The grid is logarithmically spaced in the radial direction and linearly in the azimuthal direction. This resolution was adopted for all the simulations in which the inner disk boundary has been set to $r_{\rm sc}$=1~au. However, for the three simulation with the lowest initial cloud mass (M0.15. M0.275, and M0.40), we selected an inner boundary $r_{\rm sc}=0.5$ au. Given the small $\beta$ value selected, most of the collapsing material would land inside the inner boundary condition, resulting in an unrealistically small disk, if the inner boundary were too large. This choice of $r_{\rm sc}$ is a fair compromise, considering that computation time increases for a smaller inner boundary. It should be noted that, even with this choice, the formed disks may still be slightly smaller and less massive than for a perfect simulation with the inner disk boundary extending to the stellar magnetosphere. In these three cases, we increased the number of cells in the radial direction  
and selected 323$\times$256 for M0.15, 347$\times$256 for M0.275, and 362$\times$256 for M0.40, corresponding to about four cells per scale height at half of the gas disk radius.

\subsection{Postprocessing and analysis}
\label{subs:postpr}
To be able to compute gas and dust disk properties, we need to separate the material that is contained in the disk from the one that still belongs to the envelope. 
To find this separation radius, we move from the outer computational boundary towards the star and determine the location where the radial velocity gradient  $d v_r / dr$ turns from negative (infall) to positive (expanding disk's outer edge).
In addition, we require that the disk radius delimits the area in which the gas surface density $\Sigma_{\rm g}$ exceeds 10$^{-2}$ g~cm~$^{-2}$. This latter condition becomes important in the disk evolution stages when the envelope depletes. When the disk is gravitationally unstable, its outer edge can have a complex two-dimensional shape, thus we perform an average over azimuth to obtain the disk radius.

Our simulations are post-processed to produce mock observables, allowing us to derive dust disk fluxes and masses. This enables a direct comparison with observations. The mock observables are obtained by first computing the frequency dependent absorption opacity $\kappa_{\nu}$ using \texttt{optool}~\citep{Dominik2021} and selecting three commonly used opacity prescriptions: DIANA~\citep{woitke2016}, Ricci~\citep{ricci2010} and DSHARP~\citep{birnstiel2018}. 
These prescriptions assume different grain composition and porosity and therefore lead to different estimations of $\kappa_{\nu}$ for the same dust size. We select the DIANA prescription to produce the mock observables in the main body of this paper and discuss the influence of this choice on our results in Appendix \ref{Appendix a}. The opacities are computed at the same frequency $\nu$ as the observations used for comparison, and assuming a dust distribution with ${\rm p} = 3.5$, $a_{\rm min}=5\times10^{-3}\mu m$ (consistent with the parameters adopted in the simulations; see Subsection \ref{sec:FEOSAD}), and $a_{\rm max}$ taken from the simulation output. We then compute the intensity at that given frequency in the plane-parallel limit, assuming thermal equilibrium  in the vertical direction and with negligible dust scattering:
\begin{equation}
I_{\nu}(r)=B_{\nu}(T)(1-e^{-\tau_{\nu}}),
\end{equation}
where $B_{\nu}$ is the Planck function computed at the gas temperature $T=T_{\rm g}$, assuming that gas and dust are thermally coupled\footnote{Note that FEOSAD assumes a background temperature of 15K when computing gas temperatures.}. 
The optical depth is defined as $\tau_{\nu}=\kappa_{\nu}(\Sigma_{\rm d,gr}+\Sigma_{\rm d,sm})$. 
The intensity is then integrated to obtain the corresponding total flux as: 
\begin{equation}
    F_{\nu}=\frac{1}{D^2}\int 2\pi r\ I_{\nu}(r)dr,
\end{equation}
where $D$ is the distance to the source. 
When computing the flux, we integrate only intensities above the sensitivity threshold of the survey with which we are comparing our simulations.
In this calculation we assume the disk to be axisymmetric for simplicity. However, we have tested computing the double integral over all the 2D elements in a few cases and obtained a total flux that differs by only a few percentage points.
Assuming optically thin disk emission ($\tau_{\nu}<<1$), the dust disk mass is estimated following~\citet{Hildebrand1983QJRAS..24..267H}:
\begin{equation}
    M_{\rm d}=\frac{F_{\nu}D^2}{B_{\nu}(T) \kappa_\nu},
    \label{eq:dust_mass}
\end{equation}
where $B_{\nu}(T)$ is now computed at an average dust temperature $T_{\rm d}$ and $\kappa_\nu$ is the absorption opacity calculated using $\kappa_{\nu}=2.3(\nu/230\rm{GHz})cm^2g^{-1}$. 

We computed the Cumulative Distribution Function (CDF) of observational data using the Kaplan-Meier (KM) estimator implemented in the \texttt{lifelines} package for python~\citep{Davidson-Pilon2019}, including upper limits. For the simulated disks, we assign to each a weight corresponding to the number of observed systems with similar stellar mass in the survey used for comparison. The same approach is also adopted when presenting the data distribution with violin plots.
If the survey did not provide star masses (as was the case for the observations of disks in Orion), we weight each simulated disk according to the Initial Mass Function distribution (IMF) of ~\citet{kroupa_variation_2001}. 

In this paper, we define the different Stages of young stellar objects according to the envelope mass remaining in the simulated system. We follow the results of~\citet{Dunham2010} who computed commonly used observational signatures of YSO Classes from simulations of forming disks and compared them with the remaining envelope mass. They showed that the observational boundaries between Classes is crossed at increasingly higher percentages of remaining envelope mass as the initial cloud mass increases. Following their results, we define Stage 0 as the period between the moments when 50$\%$ and 25$\%$ of the initial envelope mass remain in the simulations, for the less massive half of our sample. For the rest of the sample, we use 75$\%$ and 50$\%$. Stage I is defined  as the period during which the envelope mass decreases from 10$\%$ to 1$\%$ of its initial value, for the least massive simulations, and 25$\%$-10$\%$, for the more massive ones. 
In Stage II the evolution of the observational signatures tend to converge for different initial cloud masses. We thus define Stage II as the phase between the time when 1$\%$ of the envelope mass remains and a system age of 1~Myr (our terminal simulation time), for all simulations. 
Unless otherwise specified, the value of a quantity at a given Stage is the mean of its values at the two boundaries just defined.

\section{Results}
\label{sec:Results}
\subsection{Dependence on initial mass}
\begin{figure}
	\centering
	\includegraphics[height=1.\columnwidth]{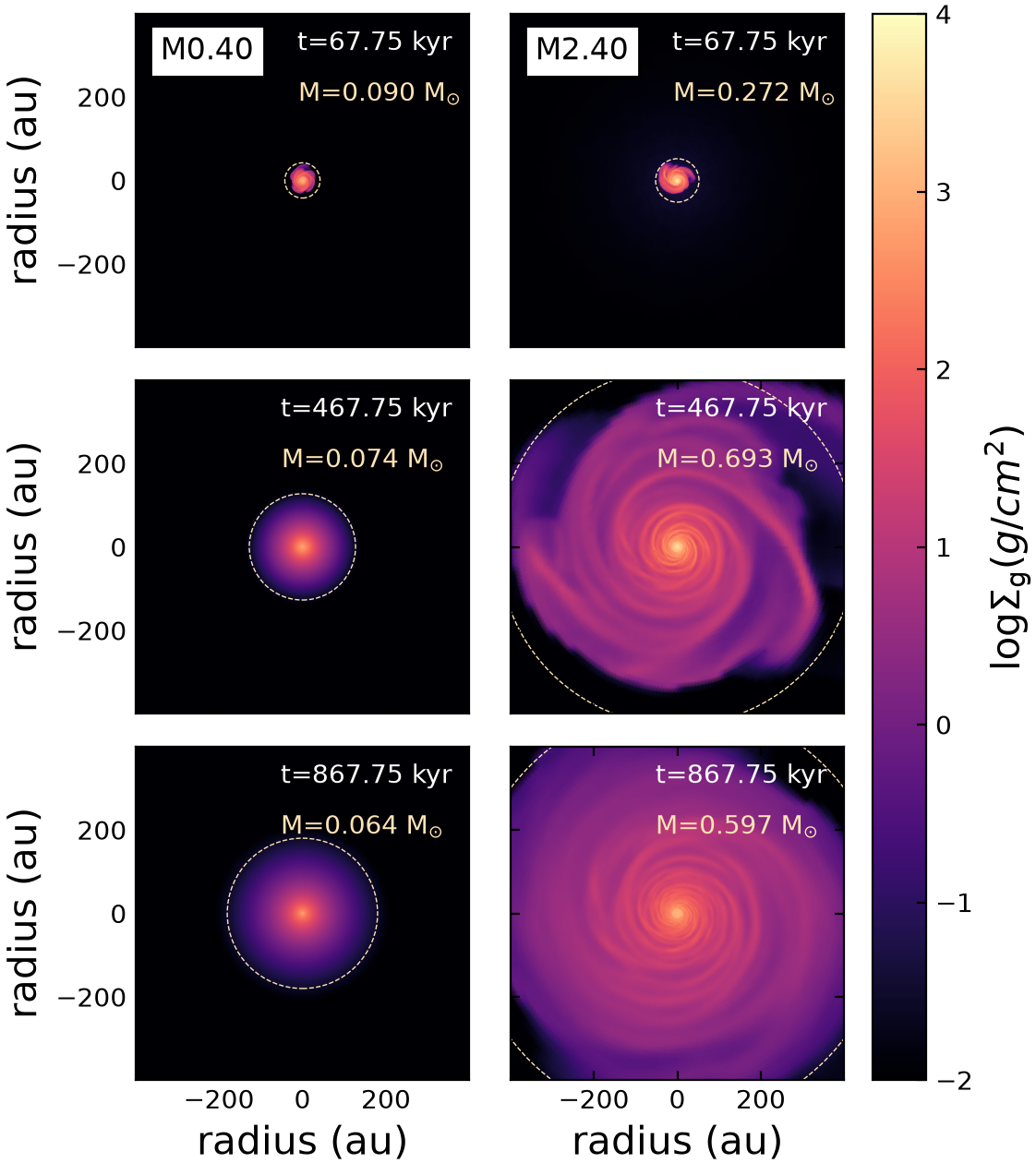}
	\caption[]{Snapshots of the gas disk surface density at different times for two simulations: M0.40 (left panels) and M2.40 (right panels). The panels also show the gas radius of each disk and the gas mass it contains. } 
	\label{fig:gas_2D}
\end{figure}
\begin{figure}
	\centering
	\includegraphics[height=1.\columnwidth]{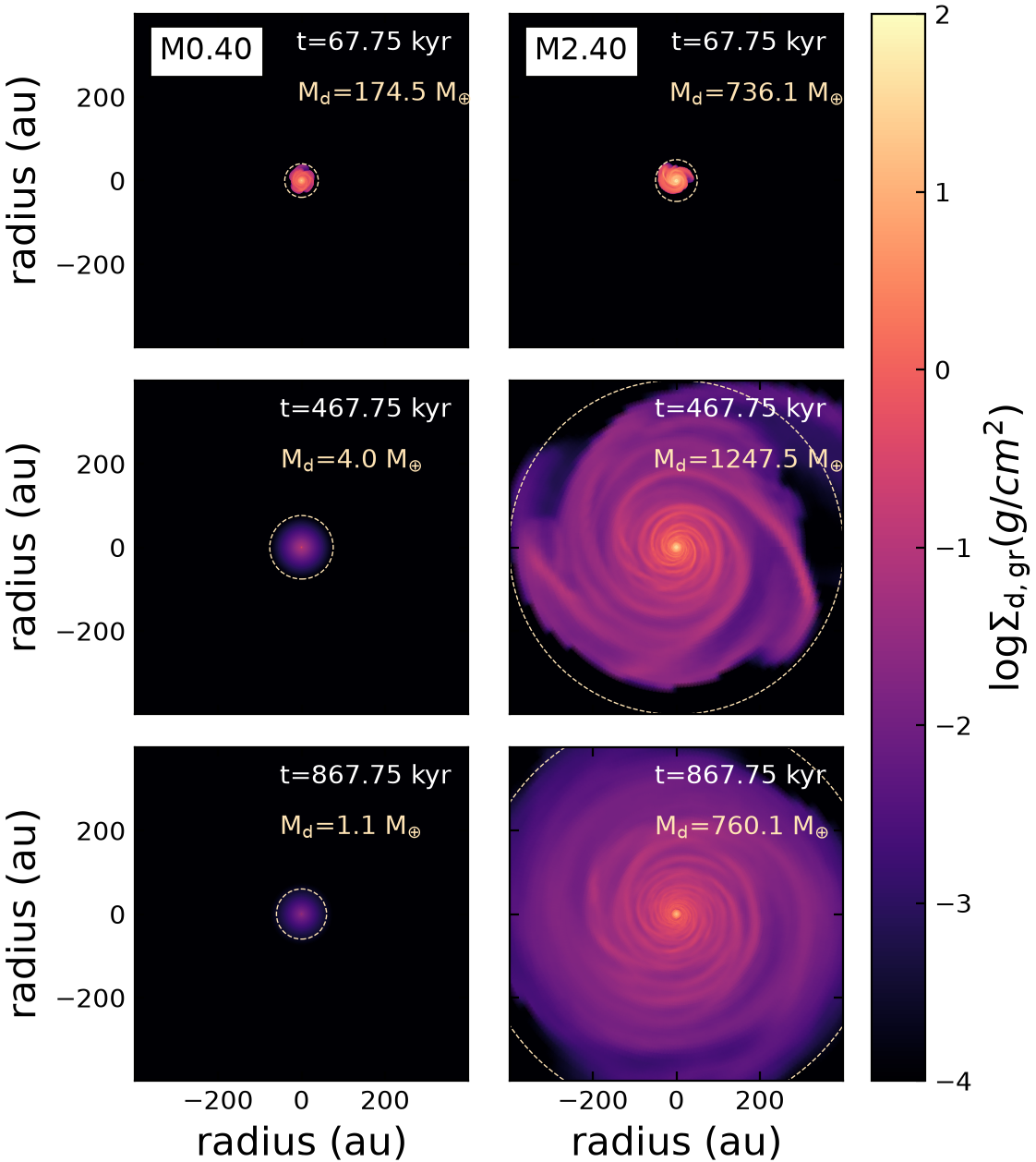}
	\caption[]{Snapshots of the dust disk surface density (for large dust) at different times for two simulations: M0.40 (left panels) and M2.40 (right panel). The panels also show the dust radius of each disk and the large dust mass it contains. }
	\label{fig:dust_2D}
\end{figure}
Figure~\ref{fig:gas_2D} shows the time evolution of the gas disk surface density for two simulations (M0.40 in the left panels and M2.40 in the right ones). The panels also show the radius of the gas disk as the separation between the disk and the envelope, as defined above, as well as the gas mass contained within it, at different times. 
Both disks form small from the infalling cloud and then spread viscously with time. 
During this time, the disks undergo a gravitationally unstable phase, with both simulations reaching a Toomre parameter value of $Q$<2 in various locations of the disk~\citep{Toomre1964}. 
However, the M2.40 simulation produces a more gravitationally unstable disk, as the initial disk-to-star mass ratio is higher due to a more massive core with the same $\beta$ value. 
The two simulations form disks with considerably different masses and radii. 
The evolutionary timescale also depends  on the initial cloud mass. After $\sim$0.5~Myr, while the M0.40 simulation has already developed into a fully Keplerian disk, the M2.40 one is still in  the gravitationally unstable phase (middle panels of Fig.~\ref{fig:gas_2D}).

Figure~\ref{fig:dust_2D} shows the time evolution of the dust disk surface density (for large dust) for the same two simulations shown in Fig.~\ref{fig:gas_2D}. 
The panels also show the radius of the dust disk, defined as the location inside which the surface density of large dust exceeds 10$^{-4}$ g cm$^{-2}$, and the dust mass contained within it
at different times.
As for the gas evolution, the masses and sizes of the dusty disks depend  on the cloud mass. The difference in evolutionary timescale between the two simulations is even more evident than for the gas disk.  In the M0.40 simulation, dust drift begins early and most of the dust disk material has already drifted towards the central star after $\sim$0.9~Myr (left panels of Fig.~\ref{fig:gas_2D}), contrary to the M2.40 simulation (right panels of Fig.~\ref{fig:gas_2D}).

\begin{figure}
	\centering
	\includegraphics[width=\columnwidth]{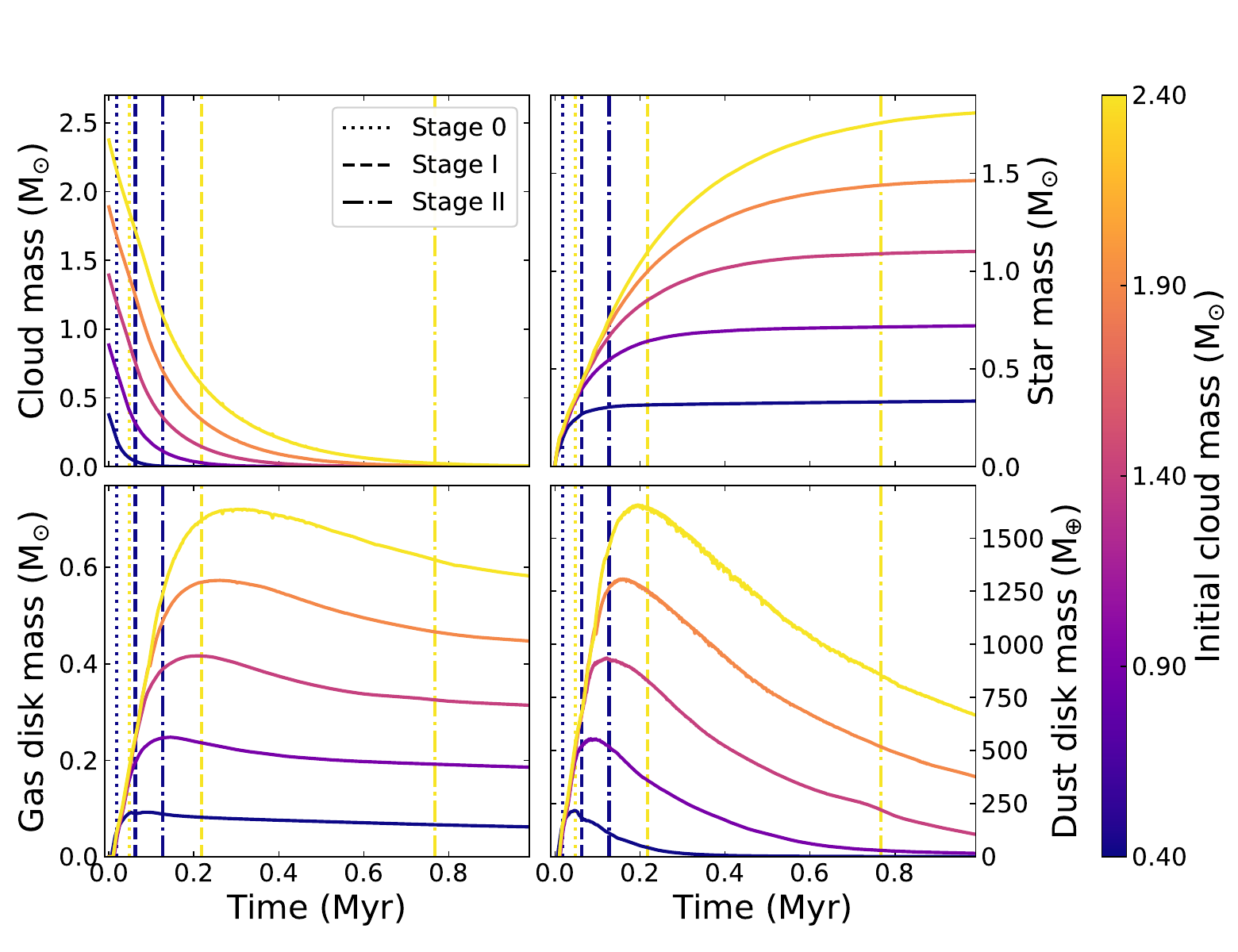}
	\caption[]{Time evolution of cloud mass (upper left), star mass (upper right), gas disk mass (lower left) and dust disk mass (lower right) for 5 simulations with different values of the initial cloud mass. The darkest color represents the least massive simulation (M0.40), while the lightest represents the most massive (M2.40). The vertical lines are showing the start of Stage 0 (dotted line), of Stage I (dashed line) and of Stage II (dash-dotted line) for two different simulation, and the color is again associated with the corresponding initial cloud mass. In particular, we show the Stage definitions for M0.40 (blue) and M2.40 (yellow).}
	\label{fig:mass_ev}
\end{figure}
\begin{figure}
	\centering
	\includegraphics[width=\columnwidth]{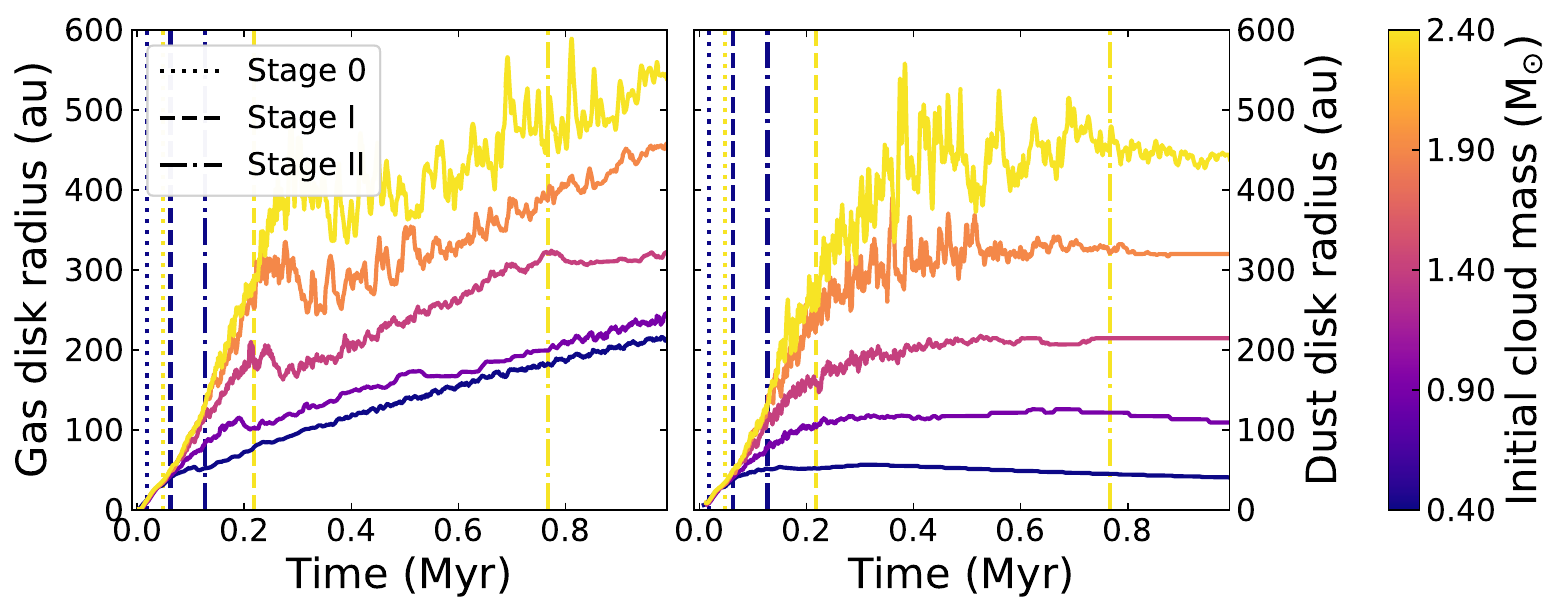}
	\caption[]{Same as Fig. \ref{fig:mass_ev} but for the gas disk radius (left panel) and the dust disk radius (right panel).}
	\label{fig:rad_ev}
\end{figure}
Figure~\ref{fig:mass_ev} shows the time evolution of the envelope, star, gas and dust disk masses for 5 simulations with different initial cloud masses. We follow the evolution of the different simulations from when the envelope starts its collapse to form a central star and a circumstellar disk. The formed stars (upper right panel) span one order of magnitude in mass, with Stage II objects ranging from M to B-type stars. Disks of very different masses also form in these simulations (lower panels). When disks initially form, the disk masses are comparable to the masses of the central stars, similar to what reported in previous works~\citep[e.g.][]{Bate2018,Lebreuilly2024}. This naturally explains the gravitational instability phase observed in all simulations~\citep{Dong2015}. The disk-to-star mass ratio reaches a value of $\sim0.3$ by the end of the infall phase, which is consistent with the results of previous hydrodynamical simulations ~\citep{Bate2018,Lebreuilly2021} and slightly higher than in MHD simulations~\citep{Lebreuilly2021,Lebreuilly2024}.

As discussed above, the different simulations evolve at very different rates. In Fig. \ref{fig:mass_ev}, the evolutionary timescale of the cloud depletion depends on the initial cloud mass. More massive clouds are also more extended 
and are characterized by lower densities in their outer regions. Thus, the free-fall time of more massive cores is longer than that of their less massive and more compact counterparts, resulting in a longer duration of the embedded stage \citep{Vorobyov2010}.
The least massive star reaches an almost constant mass after $\sim$0.1~Myr, corresponding to roughly the time at which the infall phase has finished. By contrast, the most massive star continues to grow substantially until $\sim$0.8~Myr. Disks also evolve more slowly with the increasing mass of the system. In particular, the time taken for the dust disk to dissipate depends heavily on the initial cloud mass. Only in the least massive disks the dust has almost completely dispersed by the end of the simulation. 

Figure~\ref{fig:rad_ev} shows the time evolution of the gas and dust disk radii. Changing the initial cloud mass generally results in more extended disks being formed. The gas disk radii keep increasing with time, as it is expected for viscously evolving disks~\citep{Lynden1974}.
The dust disk radii also increase with time, to then start decreasing at later times due to dust drifting towards the central star. This can be observed in the least massive simulations. For example, the gas-to-dust size ratio of the M0.40 simulations (blue curves) changes from $\sim$1 at the beginning of the simulation to $\sim$5 after 1~Myr. For the most massive simulations, instead, this ratio has only slightly increased by this time. Once again, the evolution timescale depends on the initial cloud mass.

Figure~\ref{fig:mass_ev} and Figure~\ref{fig:rad_ev} also show the times at which the different evolutionary Stages defined in Subsection~\ref{subs:postpr}
begin for two simulations: M0.40 (in blue) and M2.40 (in yellow). These Stages occur at different times due to the different evolutionary timescales of the two simulations. For example, Stage II begins at $\sim$0.13~Myr for M0.40 and at $\sim$0.75~Myr for M2.40. 
Our approach follows the actual evolution of the mass reservoir in the star-disk-envelope system, with respect to the definition of YSO stages at fixed times usually used in literature. Indeed, a typical lifetime can be associated with each YSO Class, assuming that the Class II phase lasts 2~Myr~\citep{Dunham2014,Dunham2015}. Following this approach, the Class 0 phase is expected to last around 0.17~Myr, while the Class I phase is expected to last around 0.35~Myr. However, as the typical evolutionary timescale varies significantly between our different simulations, this "fixed times" approach cannot be applied. In particular, these definitions for the different YSO phases completely fail to grasp the rapidity of evolution of the least massive simulations in our sample, as already highlighted in~\citet{Redkin2025}.

\subsection{Evolutionary trends}
\label{subs:evtrend}
\begin{figure*}
	\centering
	\includegraphics[width=\textwidth]{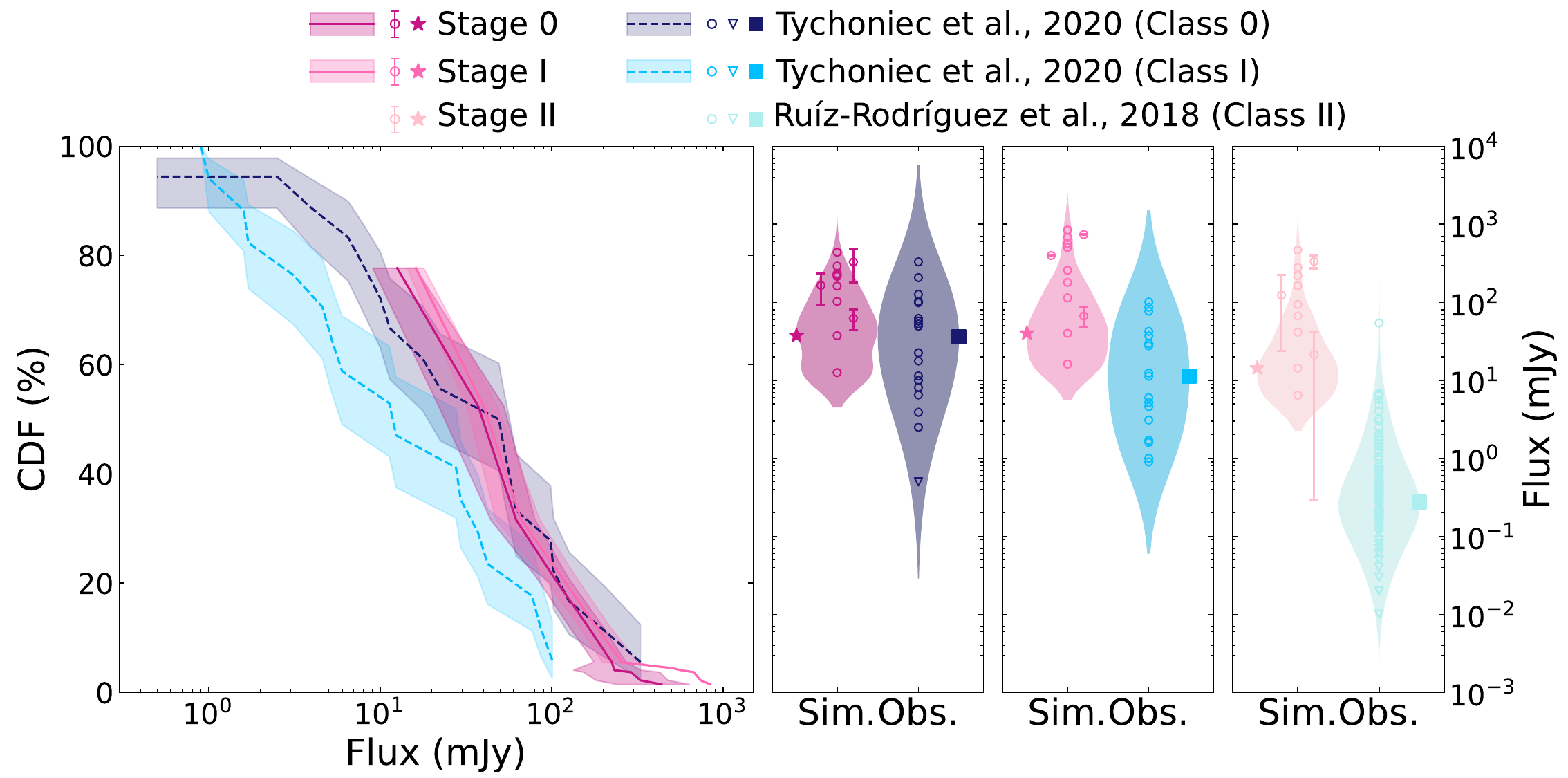}
	\caption[]{Evolutionary  trends of mm flux for disks in the Perseus star forming region vs. simulated disks. The observed fluxes have been taken from~\citet{tychoniec_dust_2020} (Class 0 and I) and~\citet{ruiz-rodriguez_alma_2018} (Class II). The colors represent the different evolutionary stages, from Class 0 to Class II, from darker to lighter. Left panel: Cumulative distribution of mm flux, with 1$\sigma$ errors shown as a shaded area for the observed disks. For the simulated disks, the shaded area shows the boundaries of our Stage 0 and Stage I definitions. Right panels: Distribution of mm flux for Class 0, Class I, and Class II disks, left to right. The circles show the data points, while the stars and squares show the medians, for the simulated and observed disks, respectively. The triangles show the upper limits of the observed disks as reported in~\citet{tychoniec_dust_2020} for the Class 0 and I objects, or corresponding to fluxes of less than three times the rms for the Class II objects. The error bars represent the boundaries of our Stage 0, Stage I and Stage II definitions for three representative data points (M0.40, M1.40 and M2.40).}
	\label{fig:Perseus_mm}
\end{figure*}

Figure~\ref{fig:Perseus_mm} presents the flux emission at 1.3 mm for disks at different stages of their evolution. From dark to light blue, we show the observed disks in the Perseus molecular cloud (Class 0 and Class I from~\citet{tychoniec_dust_2020}) and the IC 348 stellar cluster within Perseus (Class II from~\citet{ruiz-rodriguez_alma_2018}). We point out that some of the Class 0 and I sources have actually been observed at 1.1 mm and originally presented in \citep{Hsieh2019}. We exclude multiple systems with separation $<$1000~au from the sample, as reported in~\citep{Tobin2016}, since tidal interactions can cause these disks to be significantly truncated, generally lowering their continuum emission~\citep{Offner2023}.

In~\citet{tychoniec_dust_2020}, the definition of Class 0 and I is based on the bolometric temperature T$_{\rm bol}$, which is defined as the temperature of a blackbody with the same mean frequency as the source spectral energy distribution~\citep{Myers1993}. In particular, we only consider Class 0 objects that have a bolometric temperature of T$_{\rm bol}<$ 60K, and Class I objects with 90K$<$ T$_{\rm bol}<$ 650K (adapted from~\citet{Chen1995,Enoch2009}). We exclude from the sample the disks with ambiguous classification (i.e. the disks that are classified as intermediate between Class 0 and I) and those classified as Flat spectrum and Class II. ~\citet{ruiz-rodriguez_alma_2018}, instead, used the IR SED slope between 3.6 and 8~$\mu$m to classify the observed young stellar objects. In particular, they selected 128 objects where this slope has a value between -1.8 and -0.5, corresponding to Class II T-Tauri stars with optically thick disks, plus eight transitional objects that we have excluded from our analysis. The classifications chosen in both papers attempt to reflect the evolutionary state of young stellar objects, particularly in terms of whether they are embedded.  Therefore, we expect that these observed Classes agree reasonably well with our definitions of Stages(see~\citet{Dunham2010} for a discussion). 

The flux emitted by the disks in Perseus at millimetre wavelengths decreases as the disks evolve. This is usually associated with a significant amount of dust particles being removed from the disk by drift towards the central star~\citep{Rosotti2019,Pinilla2020}. Another possible explanation is that, between these stages, the dust particles have grown into aggregates (or even planetesimals or planets) that no longer emit at millimetre wavelengths~\citep{Gerbig2019,Lenz2019}.
The median value in Class I is one third of that in Class 0, while in Class II it is around two orders of magnitude lower. 

Figure~\ref{fig:Perseus_mm} also shows the results from our simulations. The mm flux has been computed using the DIANA prescription for the dust opacities (see Appendix~\ref{Appendix a} for a comparison with the other opacity prescriptions) and assuming a mean distance of 293 pc to Perseus and 310 pc to IC 348. To compute the distributions, we obtain the masses of the simulated central star when 1$\%$ of the envelope mass remains and compare them to the central star masses of the Class II disks reported in~\citet{ruiz-rodriguez_alma_2018}. Each simulation is assigned a weight representing the frequency of similar mass stars in the observed sample. We also assign weights to stars that are present in the observational sample but either more or less massive than those we simulated, in order to correctly compute the median of the distributions and CDF values. 
The same weight is used for the same simulation in all three Stages, therefore assuming a self-similar distribution of star masses throughout evolution.

The simulated disks emit a similar amount of flux in Stage 0 compared to the Class 0 Perseus disks. We are in good agreement with both the median and the upper part of the distribution. We do not reproduce disks with a flux density of less than 10 mJy, but this is probably because we do not simulate the collapse of cores with a low enough mass. This general agreement with the observations suggests that the primary agent of transport in Class 0 Perseus disks are gravitational instability and turbulent viscosity. However, the median flux of our Stage I disks is four times larger than the observed one, and the entire flux distribution of the simulated disks is shifted toward the upper end of the Perseus Class I disks. The discrepancy increases to two orders of magnitude in Stage II and the distributions of the simulated and observed disks are incompatible. This disagreement may be due to the fact that IC 348 is considered to be a 2-3~Myr old cluster~\citep{ruiz-rodriguez_alma_2018}, a typical age for intermediate Class II YSOs, while our definition of Stage II refers to younger disks. Indeed, the agreement becomes better when considering the lower limit of the error bar, corresponding to 1~Myr old disks. More generally, our results do not follow the evolutionary trend suggested by the observations, as the flux emitted by the simulations remains nearly constant in their early evolution, suggesting that dust has not yet drifted efficiently or has not yet grown sufficiently
to stop emitting at mm wavelength. 

In the simulated Stage 0 disks, the maximum Stokes number reached by dust at 10~au is typically $10^{-3}$. For simulations of intermediate mass, this corresponds to drift timescales of around 0.5~Myr, which is longer than the combined duration of Stage 0 and Stage I (cf. Fig. \ref{fig:mass_ev} or Fig. \ref{fig:rad_ev}). GI spirals may further slow down the radial drift of dust particles~\citep{Rice2004,dipierro_dust_2015}. 
At the same time, dust particles reach a maximum size of few cm only by the end of the simulations, therefore still contributing substantially to the mm flux. We note that in our models dust particles are not allowed to grow past the fragmentation barrier $a_{\rm frag}$ and no mechanisms for planetesimal formation are introduced. If present, the mass contained in these larger objects would not emit at mm wavelength. However, the low Stokes numbers reached by dust in our simulations rarely meet the conditions required to trigger planetesimal formation \citep{Li2021,gerosa2023,gerosa2024}.
The efficiency of dust drift increases over time as its Stokes number increases to reach $\rm{St}\sim10^{-2}$ at the start of Stage II. 
From this point onwards, the mm flux declines at a faster rate. When inspecting the error bars, we notice that the mm flux has decreased substantially when disks are 1~Myr old, at least in the least massive simulation. On the other hand, if dust traps were already present at the onset of Class II~\citep{Hsieh2025}, they would prevent dust from drifting efficiently~\citep{pinilla2012}. The comparison between our simulations, which do not include mechanisms to originate dust traps, and observations is therefore uncertain beyond the Class I stage, and is reported here for discussion purposes only.

In Appendix~\ref{Appendix b} we present the evolutionary trend plot for the Orion nebula~\citep{Tobin2020,vanTerwisga2022}. This plot shows similar results, suggesting that the conclusions drawn from the Perseus cloud are not limited to its specific environmental properties. In Appendix~\ref{Appendix c} we also compare our simulations with the Perseus disks observed at $\sim$9mm with VLA~\citep{tychoniec2018}. The observations at this longer wavelength trace less optically thick emission, enabling access to the inner region of disks. Once again, the comparison with our simulations yields similar results, but in this case it shows that our simulations are already slightly overestimating the mm flux of Class 0 disks.

\subsection{Dust and gas properties}
\begin{figure*}
	\centering
	\includegraphics[width=\textwidth]{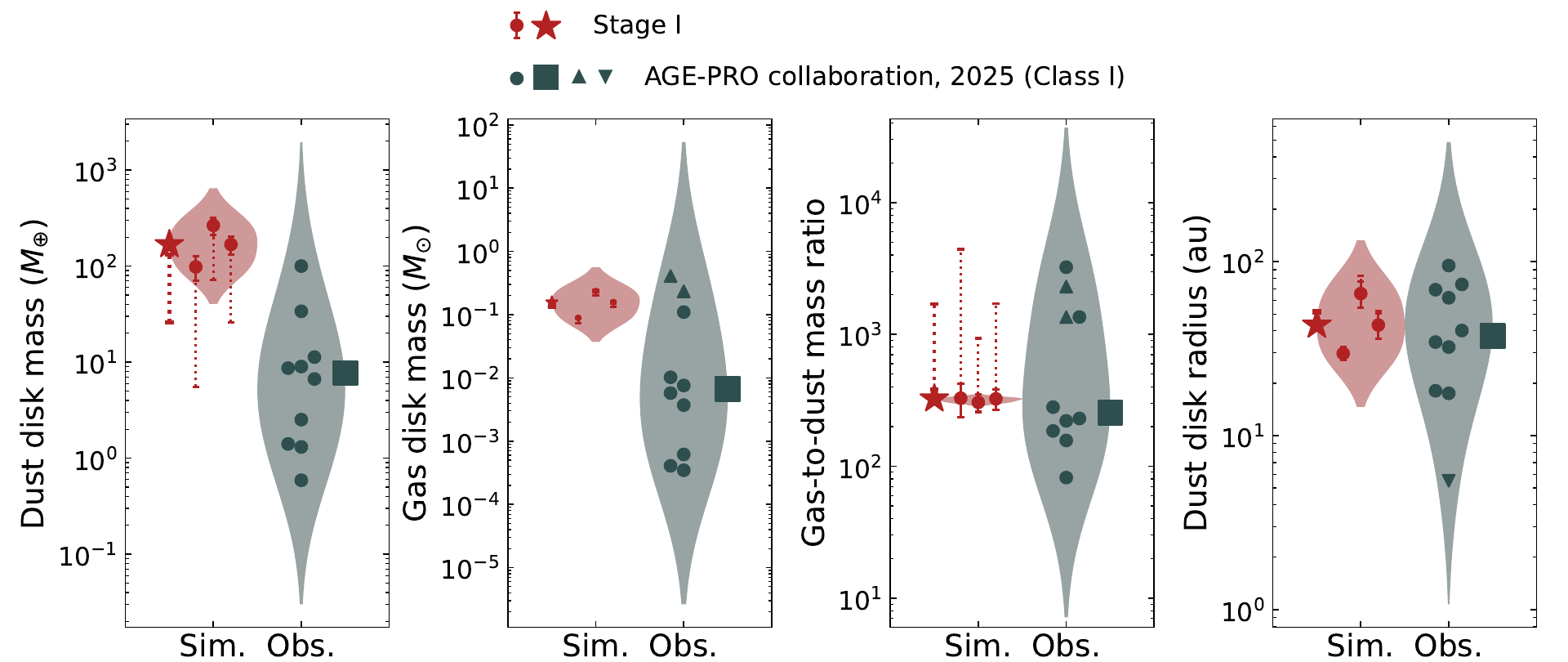}
	\caption[]{Distribution of different properties of simulated Stage I disks vs. Class I disks in the Ophiuchus star forming region. From left to right, the quantities shown are the dust disk mass, gas disk mass, gas-to-dust mass ratio and dust disk radius. The observed dust disk masses have been taken from~\citet{ruiz-rodriguez_alma_2025}, the observed gas disk masses from~\citet{trapman_alma_2025}, and the observed dust disk radii from~\citet{vioque_alma_2025}. The circles show the data points, while the stars and squares show the medians, for the simulated and observed disks, respectively. 
    The up-pointing and down-pointing arrows show the lower and upper limits, respectively, as reported in~\citet{trapman_alma_2025}. For the simulated disks, the solid errorbars (often too small to be clearly visible) represent the boundaries of the Stage I definition. The dotted error bars show the same properties computed at 0.5~Myr. In the second panel from left, the
marker size for the simulated results has been reduced so that the error bars can be noticed. }
	\label{fig:AGE-PRO}
\end{figure*}
Figure~\ref{fig:AGE-PRO} reports the dust disk and gas disk masses, the gas-to-dust mass ratio and the dust disk radius of Ophiuchus disks in the AGE-PRO sample~\citep{zhang_alma_2025,ruiz-rodriguez_alma_2025}. They selected 10 embedded disks in Ophiuchus that are considered to be younger than 1~Myr old to measure different disk properties. These disks are all classified as Class I or flat-spectrum objects, with an IR SED slope between 2 and 20~µm greater than -0.3, and a T$_{\rm bol}$ between 300 and 700~K. They also only selected disks whose central objects have a known stellar spectral type between M3 and K6, corresponding to stellar masses between 0.3~M$_{\odot}$ and 0.8~M$_{\odot}$. Finally, all the sources are without companion or with wide companions ($>$~600~au). 

Figure~\ref{fig:AGE-PRO} also presents the results from our simulations for the same four quantities. We have selected three of our twelve simulations whose central star's mass
matches the AGE-PRO selection criteria, that are M0.40, M0.65 and M0.90. To each of these simulations we assign a corresponding weight.
In particular, M0.40 represents 4 of the AGE-PRO objects, M0.65 represents 2 of them and M0.90 represents the remaining 4 objects. The values reported here as red markers have been computed for Stage I disks. However, the observed disks are considered to be late Class I. Therefore, we have added an error bar where the limiting data point is computed at 0.5~Myr. This would correspond to the end of the Class I phase following~\citet{Dunham2015} and would represent a mean age for the Ophiuchus disks in AGE-PRO, since they are believed to be younger than 1~Myr old.
However, the limit chosen is a rather extreme one, given that at this time the remaining envelope in the three simulations is 3-4 orders of magnitude less massive than their central star. Therefore, the disks of our simulations would not be embedded in their clouds at this time, while most of the AGE-PRO disks in Ophiuchus are~\citep{ruiz-rodriguez_alma_2025}. 

The left panel of Fig.~\ref{fig:AGE-PRO} shows the comparison between the dust disk mass in our simulations and the observed one. The flux from the simulated disks has been computed using the DIANA opacity prescription. The flux emitted by the observed disks has been taken from~\citet{ruiz-rodriguez_alma_2025}. For both simulations and observations, we converted the mm flux to dust mass using eq.~\ref{eq:dust_mass}. We assumed D=138.4 pc and $\kappa_{230\rm{GHz}}$=2.3\,cm$^2$g$^{-1}$, and computed $B_{\nu}$ at $T_{\rm d}=30$\,K.  
The median of the dust disk mass in our simulations is more than one order of magnitude larger than the median of AGE-PRO disks. Moreover, our disks are unable to reproduce the full two-orders-of-magnitude spread in the observed dust masses of disks, particularly the less massive part. Our simulations agree better with observations when considering the error bar at 0.5~Myr. The observed disks are believed to be optically thick. \citet{Pulgares2026}, using a multi-wavelength approach and the Ricci opacity prescription, finds that the median dust mass of Ophiuchus disks is $\sim$7 times larger than the one from~\citet{ruiz-rodriguez_alma_2025}.
Similarly, the dust mass content from our simulation has been obtained assuming optical thinness, while the simulated disks are at least partly optically thick. Indeed, the real dust mass they contain (cf. Fig. \ref{fig:mass_ev}) is almost double the dust mass computed from the flux. Nonetheless, a better estimation of the observed dust disk masses 
could help reduce the disagreement of our simulations with observations.

The second panel shows the gas disk mass estimations from simulations and observations.  The gas masses in Ophiuchus have been taken from \citet{trapman_alma_2025}. Here we adopt a gas disk definition more directly comparable to observational estimates. Since FEOSAD includes a simplified evolution of volatiles~\citep{Molyarova2025}, we are able to compute the column density of CO at each radius. We define the gas disk size $R_{\rm{CO}, 90\%}$ as the radius at which the column density $N_{\rm CO}=5\times10^{15}$~cm$^{-2}$~\citep{Zagaria2023,Trapman2023}, corresponding to the radius beyond which CO can no longer self-shield and becomes photo-dissociated~\citep{vanDishoeck1988}. During the embedded phases, however, we find that the disk and the inner envelope are too dense for this threshold to be reached. In these cases, we instead use the definition of gas disk size given in Subsection \ref{subs:postpr}, which also allows us to distinguish the disk from the envelope. The gas disk mass is then computed as the gas content within the gas disk radius. We note that different definitions of gas disk might have a substantial impact on the inferred gas disk radius, but very limited on the gas disk mass estimation because the gas density drops rapidly near the disk's outer edge. The marker size in this panel is reduced so that the errorbar is visible.
When comparing simulations and observations, we find a clear discrepancy. The median of the gas disk mass is more than an order of magnitude larger in the simulated disks with respect to the observed ones. In this case, the estimated masses for 0.5~Myr old disks are almost identical to those in Stage I, and are also in  disagreement with the observations. It is worth noting that we reproduce well the most massive disks, while we do not reproduce  the spread in masses  from observations, that is of almost 4 orders of magnitude. 

The third panel shows the gas-to-dust mass ratio. We reproduce well the median, that is slightly higher than the canonical value of 100. This could be due to the optical thickness of disks, resulting in low estimations of dust masses. However, we are unable to reproduce the lower and higher data points of the observed distribution. The latter can be matched when considering the 0.5~Myr old disks because the dust masses decrease with time, while the gas masses remain relatively unchanged, as discussed above. Our simulations would instead agree better with the lower points when considering the presence of dust traps in some Class I disks~\citep{Hsieh2025} that would stop the dust from drifting~\citep{pinilla2012}.

Finally, the last panel shows the dust disk radii. The ones from observations are computed in~\citet{vioque_alma_2025} as the radius at which 90$\%$ of the continuum flux is contained. In the same way, the dust disk radii from simulations are not the ones shown in Fig.~\ref{fig:dust_2D} and Fig.~\ref{fig:rad_ev}, but computed as the radius containing 90$\%$ of the mm flux. We reproduce the median of the observations quite well, as well as the spread (excluding the smallest observed disk). Fig.~\ref{fig:Orion_radius} compares the sizes of our simulated disks with the Class 0 and I dust disk radius of objects in Orion, showing a worse but still reasonable agreement.  
In the last panel of Fig.~\ref{fig:AGE-PRO} we also show that, up to 0.5~Myr, the dust disk sizes in our simulations are not particularly changing, while we have seen above that the dust disk masses do. This is consistent with the AGE-PRO results, which showed a decrease in dust mass between Ophiuchus and Lupus (early Class II disks), while the median dust disk radius remains almost constant between the two \citep{vioque_alma_2025}.  
Nonetheless, we find that the simulated median gas-to-dust size ratio increases from 1.7 in Stage I, to 4 at 0.5Myr. This ratio has been found to be almost constant and $\sim3$ for the disks in Lupus and Upper Sco~\citep{trapman2025b}, while the authors do not report this ratio for the objects in Ophiuchus. Gas disk sizes are not estimated in this region due to envelope contamination. Based on our simulations, we can 
speculate that the gas-to-dust size ratio 
might have been lower in Ophiuchus than in the older regions, to then reach a constant value when the disks get truncated by photoevaporation or the dust drift is halted by traps.

It is worth noting that some of the disks in Ophiuchus from the AGE-PRO sample are highly inclined (>70\textdegree). These disks may be misclassified as belonging to an earlier evolutionary Class, since the increased column density at high inclination can absorb the near-infrared emission. We can repeat the comparison by excluding these disks (Oph 1, Oph 2, Oph 9 and Oph 10), as well as Oph 8, the least embedded disk in the sample. However, since the inclined and non-embedded disks are scattered throughout the violin plot, removing them does not alter the median or reduce the observed spread in values. The conclusions from comparing our models to the AGE-PRO Ophiuchus disks are therefore  robust.

\begin{figure*}
	\centering
	\includegraphics[width=\textwidth]{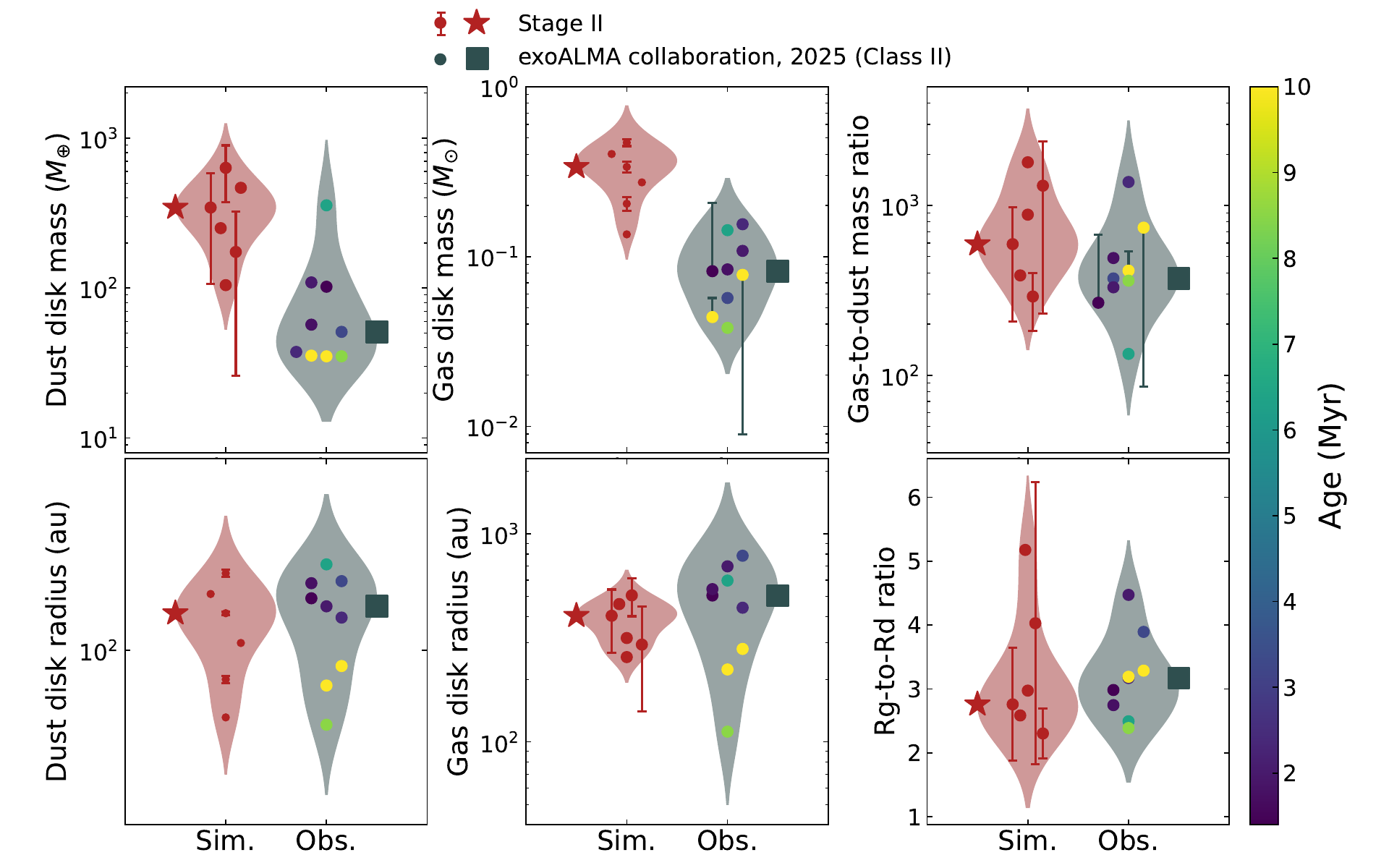}
	\caption[]{Distribution of different properties of simulated Stage II disks vs. Class II disks in the exoALMA sample. From left to right, the quantities shown are: Upper panels: dust disk mass, gas disk mass, gas-to-dust mass ratio; Lower panels: dust disk radius, gas disk radius, gas-to-dust size ratio.
    The observed dust disk masses have been taken from~\citet{Curone2025}, the observed gas disk masses from~\citet{Longarini2025}, the observed dust disk radii from~\citet{Curone2025}, and the observed gas disk radii from~\citet{Galloway-Sprietsma2025}. The circles show the data points, while the stars and squares show the medians, for the simulated and observed disks, respectively. The circles reporting the observations data are color coded by the approximate YSO age, though age estimates are highly uncertain. In the upper central and right panels, the error bars on the observed data show the estimated gas mass from~\citet{Trapman2025} and the corresponding gas-to-dust mass ratio. For the simulated disks, the error bars represent the boundaries of our Stage II definition for three representative data points.
    In the upper central and lower left panels, the marker size has been reduced so that the error bars can be noticed}. 
	\label{fig:EXO-ALMA}
\end{figure*}

The AGE-PRO collaboration studied the properties of other ten disks in Lupus and ten disks in Upper Sco. Here, we have decided to not compare our results with the objects observed in these two regions. These disks are believed to be older (1-6~Myr) and various effects that we do not consider in our simulations might become important for their longer lifespan, e.g. the presence of dust traps or external irradiation. The exoALMA collaboration targeted a sample of 15 Class II protoplanetary disks~\citep{Teague2025}. These disks are 2–10 Myr old, thus the comparison between our simulations and the exoALMA sample should be treated with caution for the same reasons mentioned above. Despite these caveats, we include this comparison because this program specifically targeted some of the brightest known disks, offering a complementary perspective to AGE-PRO.
Figure~\ref{fig:EXO-ALMA} reports the dust and gas disk masses, the gas-to-dust mass ratio, the dust and gas disk radii, and the gas-to-dust size ratio of 9 of the disks in the exoALMA sample. We selected these disks as their dynamical mass has been computed in~\citep{Longarini2025}, and excluded V4046 Sgr that is a known spectroscopic binary.

Figure~\ref{fig:EXO-ALMA} also presents the results from our simulations for the same six quantities. We have selected six of our twelve simulations whose central star's mass 
matches the star masses of the observed YSO reported in~\citep{Longarini2025}, as already done for the previous figures. The values reported here as red markers have been computed for Stage II disks. Our simulations have run up to 1~Myr, while the disks in the exoALMA sample are considered to be older. Therefore, we must be careful when drawing conclusions from this comparison.

The upper left panel of Fig.~\ref{fig:EXO-ALMA} shows the comparison between the dust disk mass in our simulations and the observed one. The flux from the simulated disks has been computed using the DIANA opacity prescription. The flux emitted by the observed disks has been taken from~\citet{Curone2025}. For both simulations and observations, we converted the mm flux to dust mass using eq.~\ref{eq:dust_mass}. We assumed $\kappa_{332\rm{GHz}}$=3.3$cm^2g^{-1}$ and used the distances reported in~\citet{Curone2025} for the observed disks, while we selected a mean distance of D=150pc for the simulated ones. We computed $B_{\nu}$ at $T_{\rm d}=20$ K. 
When considering the lower boundary of our Stage II definition, corresponding to 1~Myr old disks, the median from our simulations is only $\sim$2 times higher than the median of the disks in exoALMA. We are also able to reproduce the one-orders-of-magnitude spread in the observed dust masses of disks. The agreement gets  better when excluding the oldest disks, which lie in the lower part of the violin plot. 

The upper central panel shows the gas disk mass estimations from simulations and observations. The gas mass is computed from the simulations as the amount of gas contained within the gas disk size, defined as above.
The marker size is reduced so that the errorbar is visible. The gas masses for the exoALMA disks have been taken from \citet{Longarini2025} and, contrary to the disks Ophiuchus, they have been computed dynamically. We report as errorbars to the observed data some of the emission line-based gas masses computed in~\citep{Trapman2025}. The two estimations can disagree up to almost an order of magnitude.
The median gas mass of the simulated disks is four times higher with respect to the observed one. Once again, we find a better agreement when excluding the oldest disks. While many of the simulated disks are more massive than the observed ones, we find a one-order-of-magnitude spread in gas masses, similar to the exoALMA results.

The upper right panel shows the gas-to-dust mass ratio. The median is well reproduced. However, the simulations would increasingly overestimate the median of the observed disks with time (as shown by the upper limit of the error bars to the simulated disks, computed at 1~Myr). This is  due to the fact that, as discussed above, the gas component of the simulated disks evolves at a slower rate than the dust counterpart. The spread in values is also well reproduced.

The lower left panel shows the dust disk radii. The ones from observations are computed in \citet{Curone2025} as the radius at which 90$\%$ of the continuum flux is contained. As above, the dust disk radii from simulations are not the ones shown in Fig. \ref{fig:dust_2D} and \ref{fig:rad_ev}, but computed as the radius containing 90$\%$ of the mm flux. We reproduce the median of the observations well, as well as the spread. The marker size is reduced so that the errorbar is visible since, once again, the dust disk radii from our simulations do not change particularly with time. We have to, however, keep in mind that our disks have evolved for only 1~Myr. Their dust size might decrease for longer evolutionary times, or contrarily, might stay stable if dust traps appear in these disks~\citep{Pinilla2020}.

The lower central panel shows the gas disk radii. The ones from observations are computed in \citet{Galloway-Sprietsma2025} as the radius at which 90$\%$ of the $^{12}$CO flux is contained. 
The gas disk radius from the simulations is instead computed following the same procedure described for the second panel of Fig.~\ref{fig:AGE-PRO}.
The median of the observed gas disk sizes agrees with the median of the simulated ones. The size of our modelled disks increases with time, since they evolve under viscosity (illustrated by the upper limits of the error bars to the simulated disks). However, this does not appear to be the case for the three oldest observed disks that are smaller than the other disks in the exoALMA sample. For these disks, other mechanisms able to truncate the disk viscous expansion might be important.  

Finally, the lower right panel shows the gas-to-dust size ratio. The median of the observed disks is well reproduced. At longer evolutionary times, while the gas disk sizes increase, the simulated disks would start to overstimate the observed gas-to-dust disk size ratio if the dust disk radii remain almost constant. 

We note that we generally find a better agreement between our simulations and the properties of the disks in the exoALMA sample, with respect to the AGE-PRO disks in Ophiuchus (Fig. \ref{fig:AGE-PRO}). We can therefore speculate that the evolution of the brightest disks we observe (either the exoALMA ones or the more massive objects in Ophiuchus), at least up to few million years, might be governed by turbulent viscosity and self-gravity. On the contrary, other mechanisms not explored in our simulations might become fundamental for the early evolution of more common disks.
This interpretation is supported by a recent survey of Taurus disks at different evolutionary stages~\citep{Yesmin2026}, which found that the continuum flux of the most massive disks in their sample is similar in Class 0/I and Class II, consistent with the results of our simulations, while the median of the flux distribution decreases.
\subsection{Parameter exploration}
\label{subsec:Parameter_exploration}
\begin{figure*}
	\centering
	\includegraphics[width=\textwidth]{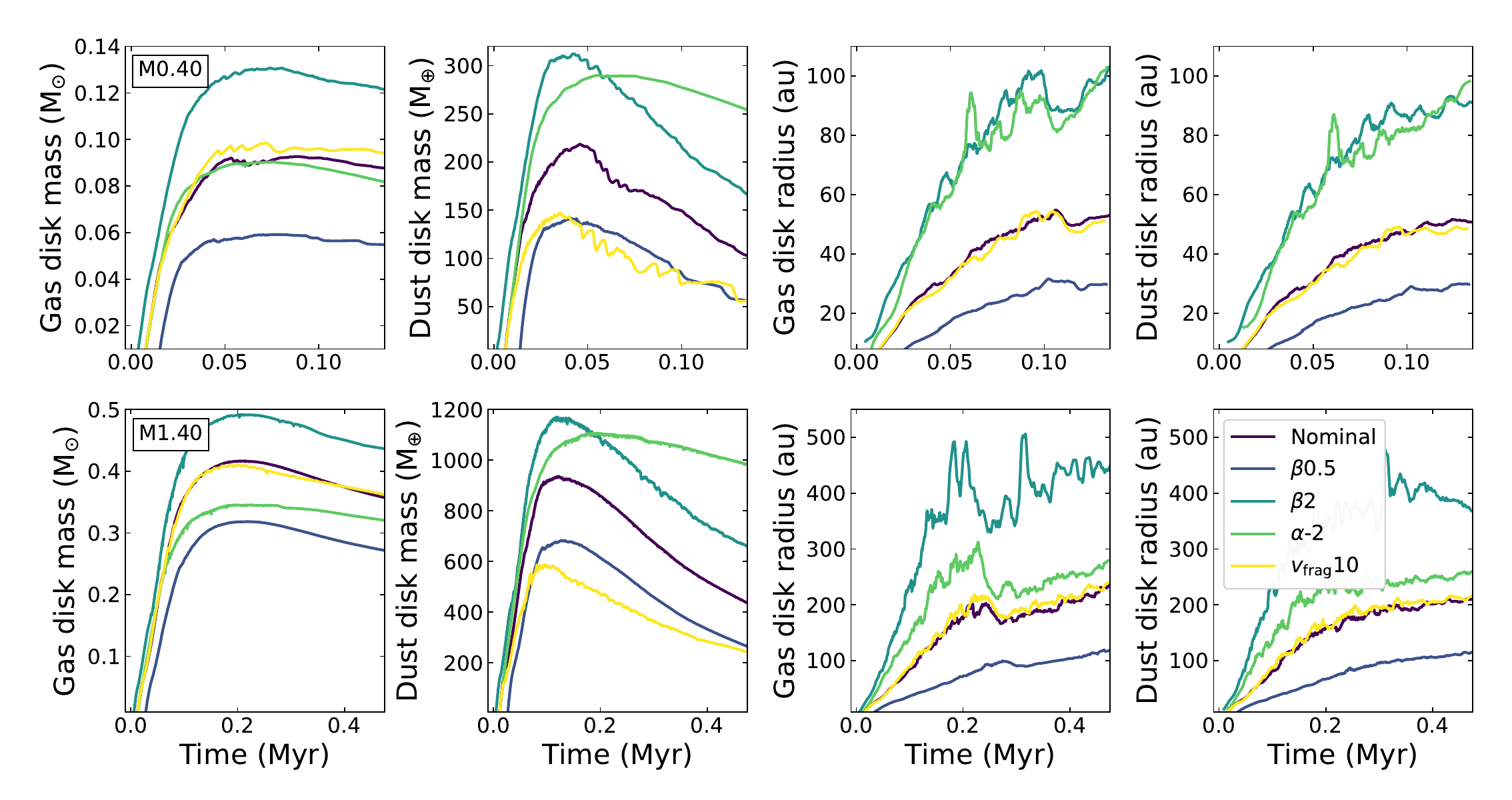}
	\caption[]{Time evolution of gas disk mass, dust disk mass, gas disk radius and dust disk radius, from left to right, for 5 simulations at different values of the $\beta$, $\alpha$, and $\varv_{\rm{frag}}$ parameters. The upper panels show the results for an initial cloud mass of 0.40~M$\odot$, while the lower panels for an initial cloud mass of 1.40~M$\odot$. The simulations at $\beta$=10$^{-3}$, $\alpha$=10$^{-3}$ and $\varv_{\rm{frag}}$=5~m~s$-1$ (nominal) are shown in dark blue, at $\beta$=5$\cdot$10$^{-4}$ in light blue, at $\beta$=2$\cdot$10$^{-3}$ in dark green, at $\alpha$=10$^{-2}$ in light green and at $\varv_{\rm{frag}}$=10~m~s$^{-1}$ in yellow. When not specified, the other parameters are kept as in the nominal simulation.}
	\label{fig:param}
\end{figure*}
We investigate the dependence of our results on three different parameters: the viscous parameter $\alpha$, the ratio of rotational-to-gravitational energy $\beta$ and the fragmentation velocity $\varv_{\rm{frag}}$. In particular, we want to understand whether the conclusions drawn from the comparisons with observations in the previous subsections are robust against this parameter exploration.

Figure~\ref{fig:param} shows the time evolution of masses and radii of the gas and dust disks for an initial cloud mass of 0.40~M$_{\odot}$ and 1.40~M$_{\odot}$ and different initial parameters, from formation up to the start of Stage II. We have varied the ratio of rotational-to-gravitational energy to a smaller value $\beta=5\cdot10^{-4}$ and a larger one $\beta=2\cdot10^{-3}$. We also selected a larger value of the viscous parameter $\alpha=10^{-2}$. We did not consider a lower value of $\alpha=10^{-4}$ because such models may require additional mass and angular momentum transport mechanisms in the form of magnetic disk winds, which is a subject for follow-up study.
However, we selected a case with a higher fragmentation velocity $\varv_{\rm{frag}}$= 10 m s$^{-1}$, which can partially mimic the effects of a lower $\alpha$ on the dust evolution since $a_{\rm{frag}}\propto\varv_{\rm{frag}}^2/\alpha$ (see eq. \ref{eq:afrag}). Increasing this parameter does not have the same impact on the gas viscous spreading or on the accreting rate that decreasing $\alpha$ would have, but it causes and increase in the typical Stokes number of the grown particles.

For both initial cloud masses, a larger or smaller $\beta$ value correspond to a more massive or less massive gas and dust disk, as well as larger or smaller radii, respectively. This is due to the fact that this parameter changes the amount of material that accretes directly onto the star compared to the amount that goes to form the disk first. 
For example, the star mass at the start of Stage II can vary from 0.97~M${_\odot}$ for $\beta=5\cdot10^{-4}$ to 1.13~M${_\odot}$ for $\beta=2\cdot10^{-3}$, for M1.40.
The simulations with $\beta=5\cdot10^{-4}$ form less massive disks (as shown in the left panels of Fig.~\ref{fig:param}). 
However, this mass reduction is not substantial enough to fully solve the disagreement between our simulations and the observed Class I and Class II disks reported in this paper.
One possibility would be to explore even lower values of $\beta$. However, the values explored here are at the lower end of the observed $\beta$ distribution, since the lowest observed extreme is $\beta=10^{-4}$~\citep{caselli_dense_2002}. Therefore, further lowering of $\beta$ would mean choosing a highly unrealistic value to be able to match the observations. Moreover, lowering $\beta$ reduces the radius of these disks with respect to the nominal simulation value, as shown in the right panels of Fig.~\ref{fig:param}. This would actually result in poorer agreement with most of the observations (see Fig.~\ref{fig:AGE-PRO},  Fig.~\ref{fig:EXO-ALMA} and Fig.~\ref{fig:Orion_radius}). 
The simulations with $\beta=2\cdot10^{-3}$ produce larger and, especially, more massive disks, thus not helping to reconcile simulations and observations of Class I and II disks. 
On the other hand, they could attenuate the discrepancy with observed radii of the Orion disks (see Fig.~\ref{fig:Orion_radius}). As mentioned above, in this paper we selected a nominal $\beta$ parameter well below the upper limit given in \citep{caselli_dense_2002}, in order to avoid fragmentation of the most massive simulated disk, and used this same value for all other core masses. Higher values could be adopted for the least massive simulations without triggering fragmentation. However, the disk-to-star mass ratio of Class I and II systems also increases from $\sim$0.3 in the nominal simulations to roughly double that in the $\beta2$ ones, approaching the maximum values of this ratio reported by the AGE-PRO and exoALMA programs. This shows that, even for the low-mass cores, an upper limit for the $\beta$ value also exists.

The simulations with $\alpha=10^{-2}$ form disks that contain either very similar (M0.40), or less (M1.40) gas content with respect to the disks in the nominal models. 
In the latter case, this could improve the agreement with observations. However, this is even less effective at reducing the disk masses than lowering $\beta$. The dust disk masses actually increase with respect to the nominal case for both initial masses, worsening the agreement with Class I and Class II observed disks. This is because a larger $\alpha$ means that particles grow less (see eq. \ref{eq:afrag}) and drift less efficiently towards the central star.
At the same time, for $\alpha=10^{-2}$ both dust and gas diffuse to larger radii with respect to the nominal case, as can be seen on the right two panels of Fig. \ref{fig:param}. In most cases, our nominal models agree reasonably with the observed disk radii, but larger dust disk sizes could reconcile our simulations with Orion disks (see Fig.~\ref{fig:Orion_radius}).

Finally, we analyze the results for $\varv_{\rm frag}$= 10 m s$^{-1}$.  As expected, the gas disk masses do not vary significantly with respect to the nominal case, since the only effect that this parameter has on the gas is due to back-reaction. The dust disk masses decrease for both M0.40 and M1.40, giving similar results to those for the $\beta0.5$ simulations. This is because the fragmentation barrier increases substantially with this larger $\varv_{\rm frag}$ value, and the larger dust particles drift faster towards the central star. Finally, the evolution of both the gas and dust disk radii is the same as in the nominal simulation, since viscous spreading is primarily controlled by the $\alpha$ value. Once again, varying the $\varv_{\rm frag}$ parameter does not seem to cause a change in the disk evolution significant enough for the simulations to agree better with the observations.

Generally, we find that the choice of initial parameters of our simulations can have an effect on the evolution of the early stages of protoplanetary disks. We also note that, in our nominal simulations, each stellar mass maps to a single disk outcome at a given Stage. The results in Fig.~\ref{fig:param} would help spreading our results.
However, in the limits tested here, the properties of these disks change only marginally (up to twice or half) when different parameters are selected. Our conclusions thus appear robust against this parameter exploration. Varying these parameters alone is unlikely to resolve the discrepancies reported above between observations and simulations, in terms of both median and spread of the distributions.

\section{Discussion}
\label{sec:discussion}

We showed in the previous section that the results of our simulations can reproduce the observed dust fluxes of Class 0 disks. This result suggests that in the very early stages of disk formation and evolution, while the molecular cloud is still infalling onto the disk, self-gravity acts as the primary transport mechanism, and viscosity as secondary, in agreement with the strength of the corresponding torques in the disk \citep{Redkin2026}. The choice of initial parameters can have an effect on this early stage, as shown in Fig. \ref{fig:param}, but they would only marginally change the properties of the formed disks. We also showed that the simulated disks are comparable in size to the Class I and Class II observed disks. However, the other quantities computed from the simulations disagree with the observed ones for Class I and Class II disks. In particular the medians of gas and dust masses are always larger than the corresponding observed values. While we can reasonably reproduce the brightest disks, we fail to replicate the low-mass end of the observed distributions. The comparison with the AGE-PRO sample also reveals that our simulations do not explain the mass spread reported in the observations.
We infer that our simulations miss a mechanism able to accelerate the disk evolution in terms of gas removal and early dust drift, while not altering substantially the disk radius. We discuss some possible solutions to test in future models. 

The finding that the gas and dust content of our disks match well with some of the most massive Class I and Class II disks in the surveys analyzed here has interesting implications. The massive disks in our simulations retain a large amount of dust (up to $\sim$500 M$_\oplus$ at 1~Myr), even in the absence of dust traps. These disks would thus naturally attenuate the "mass budget problem" of planet formation. 
Therefore, it is plausible to assume that the most massive observed disks, if evolving under gravity and viscosity only, would be retaining a large amount of dust down to the Class II phase, as occurs in our simulations, allowing for massive planetary cores to form.   

Finally, we note that the cloud collapse model used in these simulations is quite simplified. It assumes that disks form in isolation (as discussed below) and from an axisymmetric collapse, and does not account for magnetic fields or core-scale turbulence. All these physical processes are known to play an important role in shaping the properties of forming disks~\citep{Allen2003,Duffin2009,Tsukamoto2013,Kuznetsova2022}. 
This simplified approach allowed us to perform a systematic study of disk formation and evolution across a range of cloud masses, following disks up to 1~Myr and comparing them with observations at all evolutionary stages, that would not be feasible with more complex collapse models. Nevertheless, our results are broadly consistent with those of more sophisticated cloud collapse simulations, which similarly form objects with comparable dust disk size but larger dust disk mass than the observed ones (see Fig. 14 of ~\citet{Lebreuilly2024} and the comparison with~\citet{Bate2018} reported in~\citet{Tobin2020}). 

\subsection{Possible solutions to the disagreement in dust disk mass}
Different possibilities might help lower the amount of dust in the young simulated disks. Dust in our simulations is only allowed to grow up to the fragmentation barrier. However, some of these particles might grow into planetesimals or even planets in early stages of the disk evolution, and recent observations suggest an early onset of planet formation~\citep{Shoshi2025,Hsieh2025}. These larger dust aggregates do not emit at mm wavelength, reducing the continuum flux. However, the dust particles in our simulations have only reached Stokes numbers of $10^{-2}$ in Stage II and would rarely form planetesimals \citep{Li2021,gerosa2023,gerosa2024}. On the other hand, our simulations do not have any additional mechanism that could form structures (e.g. dead zones~\citep{pinilla2016}, large scale vortices~\citep{meheut2012}, snow lines~\citep{drazkowska2017}) able to trap dust particles and allow them to grow~\citep{pinilla2012}, apart from the presence of short-lived spirals due to gravitational instability. 
Of the parameters tested in Fig.~\ref{fig:param}, reducing $\beta$ or increasing $\varv_{\rm frag}$ lead to lower dust masses, but is not enough to solve the disagreement. A smaller $\alpha$ value could also help reducing the dust disk mass, causing the opposite effect of the $\alpha$-2 simulations. However, this would in turn increase the amount of gas present in the simulated disks, worsening the disagreement with observed gas masses. Finally, our calculations of observable quantities assumed face-on disks, whereas AGE-PRO and exoALMA report inclinations from 32$^\circ$ up to 74.5$^\circ$ \citep{Curone2025,ruiz-rodriguez_alma_2025}. Adopting this maximum inclination lowers the estimated dust disk masses by 30$\%$ in Stage I and 20$\%$ in Stage II. This is a significant change, but it does not improve either the agreement with the median or our ability to reproduce the spread in dust disk masses in Fig.~\ref{fig:AGE-PRO}.

It is worth noting that the observations themselves might hold an important caveat. Indeed, disks are believe to be optically thick, expecially in the early stages of their evolutions, when disks are massive and embedded in their envelope. Therefore, the observed flux and the dust mass estimation should be considered as lower limits. An evidence of this can be found in~\citeauthor{Pulgares2026}, where the authors found that the median of Ophiuchus dust disk masses increases of almost an order of magnitude when considering the more complex multi-wavelength approach. The disks in our simulations are also optically thick. However, we are able to retrieve the dust disk masses quite well (the actual dust disk masses are up to twice as large as the ones computed from the mm flux) and therefore we have to consider this difference when comparing them to the observed ones. However, we note that the disagreement actually increases when comparing our simulations with observations at longer wavelengths that trace less optically thick emission and should therefore lead to better flux and mass estimations (Fig. \ref{fig:Pers_VLA}).
Other observational caveats include the fact that the observations chosen for comparison have not been corrected for free-free emission (with the exception of those reported in Appendix \ref{Appendix c}), and that the dust disc masses derived from these observations were computed neglecting scattering. We recomputed the emergent intensity with scattering included in a subset of our models (following~\citet{Sierra2020}). We found differences of $\sim$5$\%$ on the resulting mm fluxes, which is a negligible correction compared to the uncertainty introduced by the choice of dust opacity prescription (see Appendix \ref{Appendix a}).

\subsection{Possible solutions to the disagreement in gas disk mass}
To reconcile our simulations with observations, a better approach would be to introduce a mechanism that can remove a substantial amount of gas from the disk over short timescales. This would in turn accelerate the drift of dust by increasing its Stokes number. As mentioned above, our simulations are able to reproduce some of the brightest disks in the AGE-PRO and exoALMA sample. We therefore favor processes that could explain the 
simultaneous formation of a wide range of masses and therefore helping not only to reduce the median but also to increase the spread of the distribution. 
We consider different possibilities here. In particular, we focus on the fact that our simulations consider one isolated molecular cloud forming one disk, thus completely ignoring the possible impact of its interaction with the environment.  

It is worth noting that the estimations of gas disk masses from observations also have some degree of uncertainty. They rely mostly on the assumed abundance of the tracer molecule relative to H${_2}$, as well as on the assumption that the observed lines are optically thin. In~\citet{Trapman2025} the authors compare molecular and kinematic gas disk masses and found most of them to agree within a factor of $\sim$3 (see Fig.~\ref{fig:EXO-ALMA}).

\paragraph*{Photoevaporation} One possibility would be  mass removal by photoevaporative winds, either from the the central star or from nearby massive stars. Theoretical and observational estimates of the mass loss rate due to photoevaporation span several orders of magnitude, ranging from 10$^{-10}$~M$_{\odot}$~yr$^{-1}$ up to almost 10$^{-5}$~M$_{\odot}$~yr$^{-1}$~\citep{Picogna2019,Sellek2024,Aru2024,Anania2025} depending on the considered region, photoevaporation model, star and disk properties or driving radiation. At the highest end of this range, external photoevaporation could significantly reduce disk masses.  However, in the star forming regions considered in this paper (Perseus, Ophiuchus and even Orion, see Appendix \ref{Appendix b}) external photoevaporation is expected to be largely negligible~\citep{Anania2025b}. In this context, disk mass loss through external photoevaporation would operate on timescales of millions of years and would also lead to a significant decrease in disk size\citep{Anania2025}. Internal photoevaporation might be otherwise relevant for late dispersal of disks, but less relevant for our simulations of disks in early stages~\citep{Alexander2006}.

\paragraph*{Outbursts} Outbursts are another mechanism that can substantially decrease the mass of a disk, especially around low mass stars, because up to 30\% of the final stellar mass may be accreted during the outbursts \citep{DunhamVorobyov2012}. Observational and theoretical studies suggest that protostellar evolution can undergo violent bursts of accretion, known as FU Orionis-type eruptions. These outbursts are thought to be triggered by either disk instabilities or external environmental factors, and can cause the accretion rate to increase by several orders of magnitude, reaching values as high as 10$^{-4}$~M$_{\odot}$~yr$^{-1}$~\citep{Hartmann1996,Armitage2001,Audard2014}. 
This phenomenon is especially significant during the early stages of stellar evolution, when material is still infalling from the envelope onto the disk \citep{Quanz2007,VorobyovBasu2015}, and may therefore be highly relevant to our work. However, observations of outburst systems indicate that their dust disks are on average more massive and more compact than those of quiescent systems \citep{Kospal2021}, making the relationship between outbursts and disk masses more complicated than might first appear.

\paragraph*{MHD winds} In recent years, a new paradigm involving MHD winds has been proposed for removing angular momentum from the disk and therefore allowing accretion of material to the central star~\citep{Blandford1982,Wardle1993,Pascucci2023}. Various observational~\citep{Fang2018,deValon2020,Tabone2020} and theoretical studies~\citep{bethune2017,bai2017,Lesur2021} report disk mass loss rates due to MHD winds (through both accretion onto the star and ejection) that are similar in magnitude to, or larger than, the mass accretion rate driven by viscous evolution. Therefore, simulations in the MHD winds paradigm could produce disks that lose their mass more quickly~\citep{Kadam2025,Redkin2026} and better agree with observations of Class II disks~\citep{Somigliana2023,Tabone2025}. However, it is unclear whether this difference would be evident at the earlier stages explored in this paper. At the same time, simulations including MHD winds show that the dust disk size decreases faster than in the viscous case, which could cause potential disagreement with the observational results~\citep{Zagaria2022MNRAS.514.1088Z,Kadam2025,Redkin2026}.

\paragraph*{Encounters} 
\citet{Bate2012} performed a radiation hydrodynamic simulation of star cluster formation and found that, after 10$^5$ yr, half of the stars had been involved in encounters closer than 10~au. This stochastic process would therefore affect only half of the disks in the cluster, while leaving the evolution of the other half unchanged. The encounters strip materials from the disks surrounding the forming stars and truncate them. However, the author argues that this does not mean that most disks are small, since stars may accrete a new disk from the surrounding cloud even after an encounter~\citep{Bate2003}. By the end of the star cluster simulation, the median of disk radii is around 60~au, while many disks have masses that exceeds $0.3$~M$\odot$ or lower than 10$^{-2}$~M$\odot$~\citep{Bate2018}. Therefore, including encounters in our simulations might help us to retain massive disks while also producing less massive ones, which would cause a significant spread in our results, similar to that observed in the disk surveys considered in this paper. As observational evidence of encounters has only been found in a few disks~\citep{Cuello2023}, it remains unclear what role they play in the evolution of disk populations.

\section{Conclusions}
\label{sec:conclusion}
In this paper we study the formation and evolution of protoplanetary disks, starting from the collapse of molecular clouds of different masses. These disks evolve under self-gravity and viscosity through Stage 0 and Stage I and into early Stage II. We compare these simulations with surveys of disks in different evolutionary stages to draw conclusions about the processes governing the evolution of protoplanetary disks. 

The main findings are:
\begin{itemize}
    \item Clouds of increasing masses collapse to form disks of increasing mass and radius. Less massive disks evolve faster; the gas disk becomes fully Keplerian and they deplete most of the dust content in less than one million year. On the contrary, massive disks still show spirals caused by gravitational instability and retain a considerable amount of dust by the end of the simulation.
    \item The dust content in Stage 0 simulated disks generally agrees with the observed one. The results are thus consistent with a scenario in which self-gravity is the primary transport agent, with viscosity playing a secondary role. However, the later dust evolution in simulated disks appear to be slower than in the observed ones. This suggests the need for a mechanism to increase the Stokes number and/or the size of dust particles.
    \item While the size of the Class I and Class II disks is similar to that of the simulated disks, their dust and gas content do not match. In particular, the gas disk mass of the simulated disks is significantly higher than that of the observed ones. While our models reproduce the most massive observed disks, they fail to explain the least massive ones.
    This suggests that the reported discrepancies between observations and results might be due to a missing mechanism that efficiently removes gas from some of the modeled disks, already from the early stages of disk evolution. We speculate that the most massive disks observed may have evolved through turbulent viscosity and self-gravity alone, whereas the least massive ones underwent a more complex evolution. 
    \end{itemize}
We have also performed a parameter study, finding that our results are not significantly affected by the choice of initial parameters. 
We conclude that additional mechanisms arising from the interaction of protoplanetary disks with the environment in which they form may be needed to resolve the reported discrepancies with observations, and leave their implementation to future work.

\section*{Acknowledgements}

The authors are thankful to the anonymous referee for constructive comments, which helped to improve the manuscript.
F.A.G. and P.P.
acknowledge funding from the UK Research and Innovation (UKRI)
under the UK government’s Horizon Europe funding guarantee from
ERC (under grant agreement No 101076489). 

Simulations were performed in part using Austrian Scientific Computing (ASC) infrastructure (\href{https://asc.ac.at/}{https://asc.ac.at/}). 
The authors acknowledge the use of the UCL Myriad High Performance Computing Facility (Myriad@UCL), and associated support services, in the completion of this work.

\section*{Data Availability}
The datasets generated and analyzed during the current study are available from the corresponding author upon request.



\bibliographystyle{mnras}
\bibliography{main} 




\appendix
\section{Different opacity prescriptions}
\label{Appendix a}
\begin{figure}
	\centering
	\includegraphics[width=0.8\columnwidth]{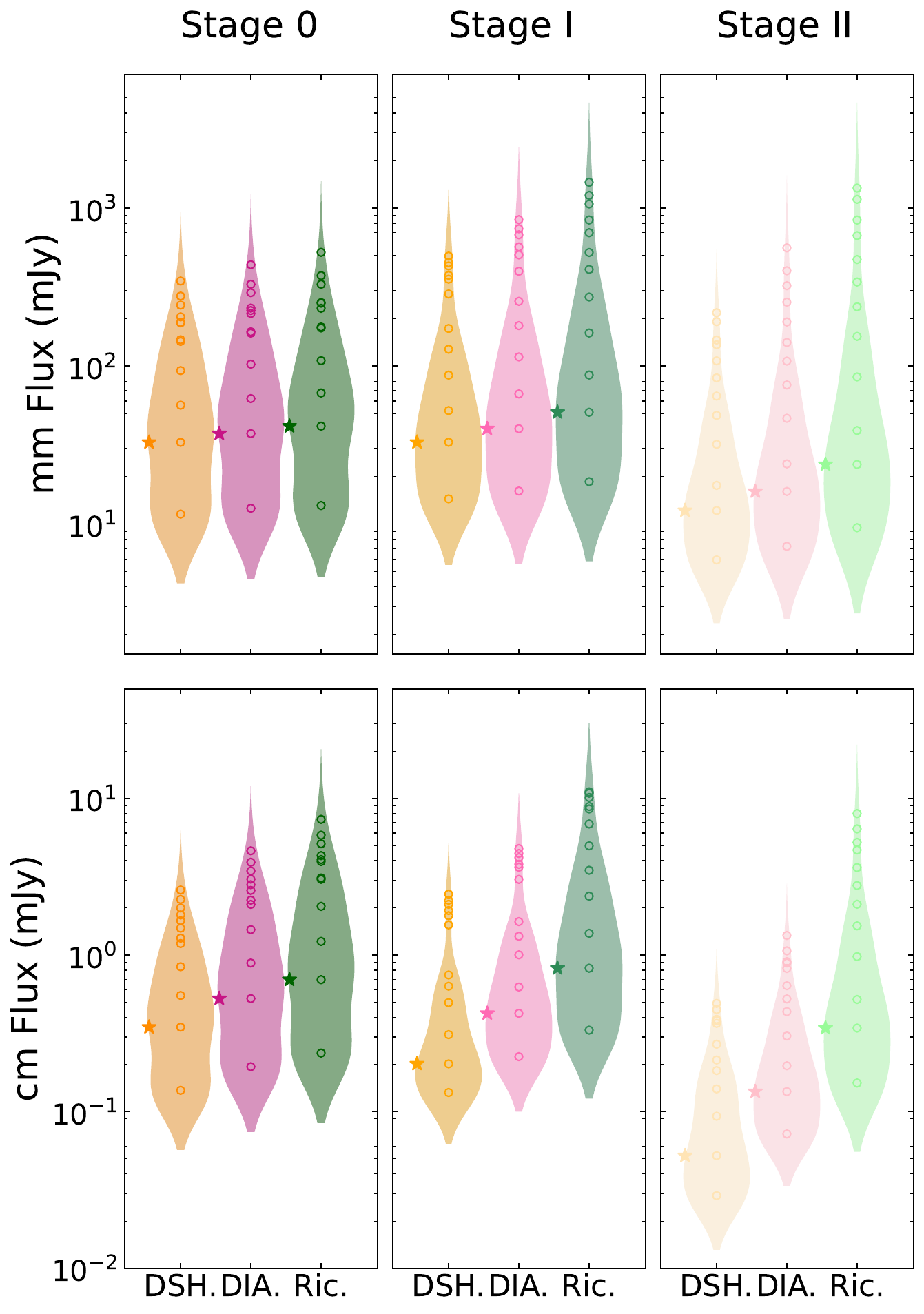}
	\caption[]{Distribution of mm (upper panels) and cm (lower panels) flux for Stage 0 (left panels), Stage I (center panels), and Stage II disks (right panels). In each panel, we show the distribution using three different opacity prescriptions: DSHARP~\citep{birnstiel2018}, DIANA~\citep{woitke2016} and Ricci~\citep{ricci2010} (left to right). }
	\label{fig:Opacities}
\end{figure}

Figure~\ref{fig:Opacities} shows the mm and cm flux of the simulated disks, as presented in Fig.~\ref{fig:Perseus_mm} and Fig.~\ref{fig:Pers_VLA}, but for three different opacity prescriptions.
In addition to the DIANA prescription~\citep{woitke2016}, we use also the ones presented in~\citet{ricci2010} (Ricci) and in~\citet{birnstiel2018} (DSHARP).
These prescriptions assume different grain composition and porosity, therefore leading to different absorption opacity values for the same dust size. For example, DSHARP and Ricci opacities include water ice in the grain composition, while DIANA does not. DSHAP considers compact dust particles, while DIANA and RICCI consider a certain level of porosity. The choice of dust opacity prescription remains one of the major sources of uncertainty in the derivation of disk dust masses, and there is currently no consensus on which prescription best represents the properties of dust in protoplanetary disks.

Figure ~\ref{fig:Opacities} shows that the mm fluxes obtained using the DSHARP prescription are always lower than when using the DIANA one, while the opposite happens when using Ricci. 
In the left and central upper panels,
the median flux is similar for the three cases. The differences between the various prescriptions become more evident for the calculations at cm wavelength (lower panels) and for Stage II disks (right panels), where the median value increases up to a factor of seven. The three prescriptions assign different weights to the contribution of large grains to the opacity and, as the maximum grain size increases in Stage II disks, this becomes more noticeable. 
Using the DSHARP prescription also gives the impression of faster evolution between stages, while the Ricci prescription leads to a slower one. However, comparing with the disks in Perseus (cf. Fig.~\ref{fig:Perseus_mm}), we are not able to solve the disagreement with the observed Class I and Class II disks even using the DSHARP prescription. Conversely, using the DSHARP prescription would improve the agreement between the observed and simulated Class 0 disks in Fig.~\ref{fig:Pers_VLA}, while still showing discrepancies with the observed Class I disks. Different prescriptions may be more appropriate to use for different wavelengths, evolutionary stages or disk properties. 

In the main text, we have decided to use the DIANA prescription as the nominal one, since it gives intermediate results. Moreover, the dust material density resulting from the grain composition selected in the DIANA prescription ($\rho_{\rm s}=$2.08$g cm^{-3}$) is the closest to the value that we use in our models.

\section{Orion}
\label{Appendix b}
\begin{figure}
	\centering
	\includegraphics[width=\columnwidth]{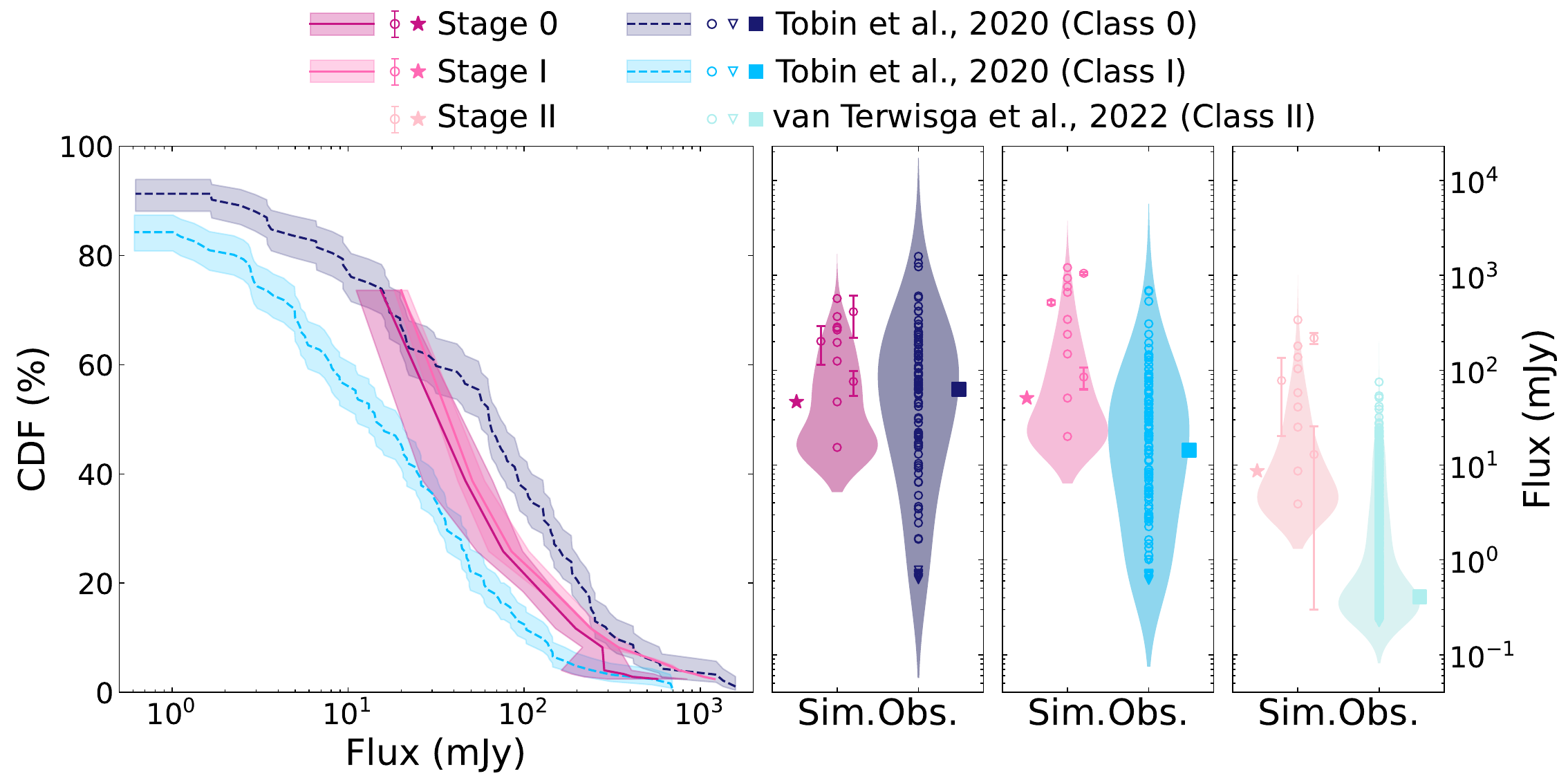}
	\caption[]{Evolutionary trends of mm flux for disks in the Orion molecular clouds vs. simulated disks. The observed fluxes have been taken from~\citet{Tobin2020} (Class 0 and I) and~\citet{vanTerwisga2022} (Class II). The colors represent the different evolutionary stages, from Class 0 to Class II, from darker to lighter. Left panel: Cumulative distribution of mm flux, with 1$\sigma$ errors shown as a shaded area for the observed disks. For the simulated disks, the shaded area shows the boundaries of our Stage 0 and Stage I definitions. Right panels: Distribution of mm flux for Class 0, Class I, and Class II disks, left to right. The circles show the data points, while the stars and squares show the medians, for the simulated and observed disks, respectively. The triangles show the upper limits of the observed disks as reported in~\citet{Tobin2020} for the Class 0 and I objects and as reported in~\citet{vanTerwisga2022} for the Class II objects. The errorbars represent the boundaries of our Stage 0, Stage I and Stage II definitions for three representative data points (M0.40, M1.40 and M2.40).
 }
	\label{fig:Orion}
\end{figure}

\begin{figure}
	\centering
	\includegraphics[width=\columnwidth]{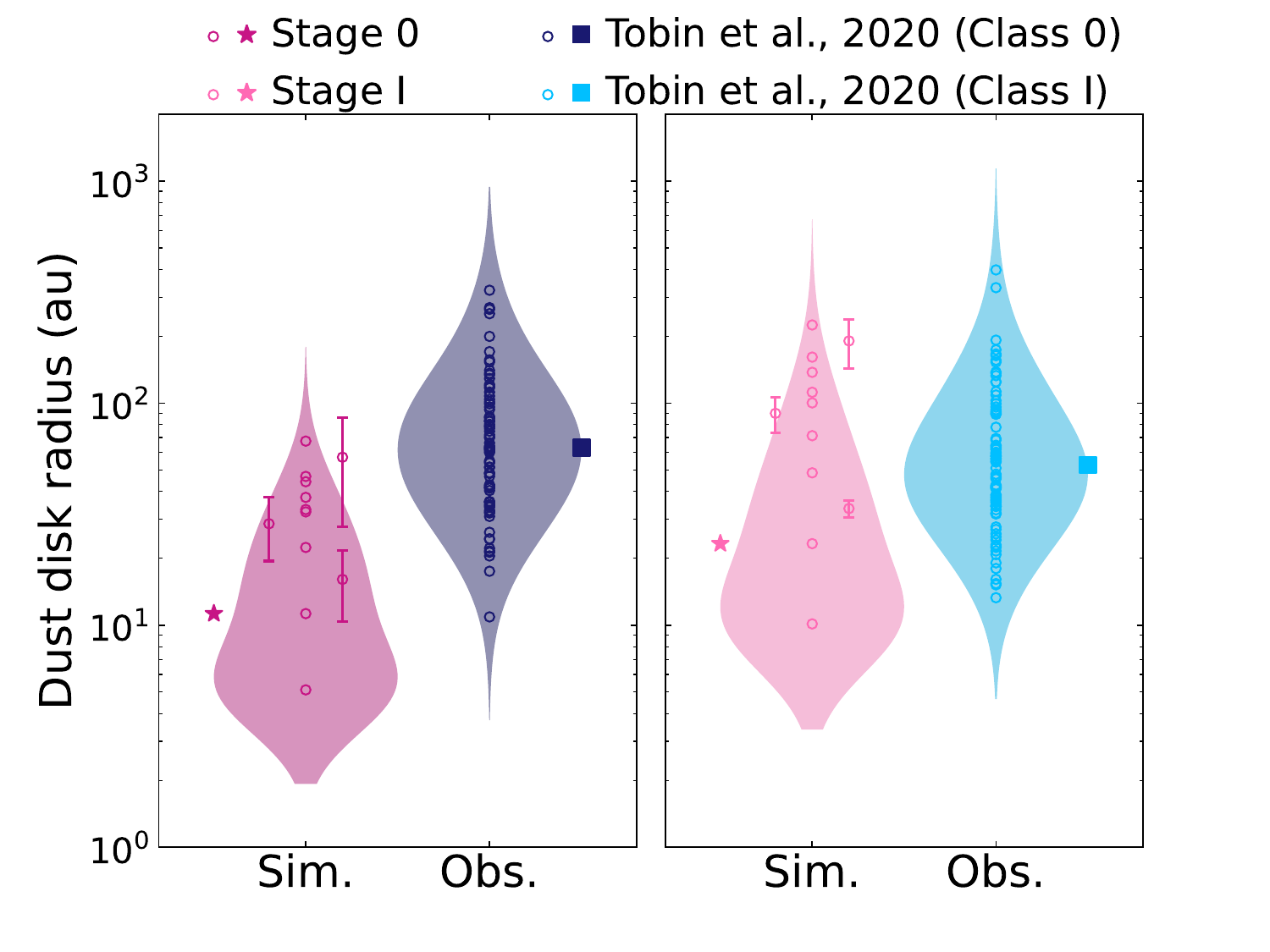}
	\caption[]{
    Distribution of dust disk radius for disks in the Orion molecular clouds vs. simulated disks in 
    Class 0 (left) and Class I (right). The observed dust disk sizes have been taken from~\citet{Tobin2020}. The circles show the data points, while the stars and squares show the medians, for the simulated and observed disks, respectively. The errorbars represent the boundaries of our Stage 0 and Stage I definitions for three representative data points (M0.40, M1.40 and M2.40).
 }
	\label{fig:Orion_radius}
\end{figure}
Figure~\ref{fig:Orion} presents the flux emission at mm wavelength for disks at different stages of their evolution. From dark to light blue, we show the observed disks in the Orion A and B molecular clouds (Class 0 and Class I from~\citet{Tobin2020} and Class II from~\citet{Tobin2020}, the latter for Orion A only). The Class 0 and I sources have been observed at 0.87 mm, while the Class II ones at 1.3 mm. We exclude multiple systems with separation
<1000 au from the sample, as reported in~\citet{tobin2022}.

In~\citet{Tobin2020}, the definition of Class 0 and I is based on the bolometric temperature T$_{\rm bol}$, that is defined as the temperature of a blackbody with the same mean frequency as the source spectral energy distribution~\citep{Myers1993}. In particular, Class 0 objects have bolometric temperature T$_{\rm bol}<$ 70K, while Class I objects have 70K$<$ T$_{\rm bol}<$ 650K (adapted from~\citet{Chen1995,Enoch2009}). We have excluded Flat Spectrum sources from the sample. We cannot rule out that some older disks may have been misclassified as earlier classes due to their high inclination, as suggested by \citet{Sheehan2022}.
~\citet{vanTerwisga2022} used an improved version of the infrared color criteria of~\citet{Megeath2012}. They selected 873 Class II disks, plus 40 less evolved objects that we have excluded from our analysis.

The presence of O-type stars in Orion is known to drive the efficient photoevaporation of material from the disks in the Trapezium region of the Orion Nebula Cluster and the NGC 2024 nebula, resulting in substantial mass loss~\citep{Mann2014,Haworth2021,Anania2025}. Therefore, it may be difficult to compare observations of disks in Orion with our simulations, in which the disks have been modeled without external photoevaporation. However,~\citet{Tobin2020} targeted only few protostars located within Trapezium and NGC 2024, while~\citet{vanTerwisga2022} selected objects at least 4 pc away from the Trapezium bright stars. Therefore, external irradiation should not have had a significant impact on the evolution of the observed disks.

Figure~\ref{fig:Orion} also shows the results from our simulations. The mm flux has been computed using the DIANA prescription for the dust opacities and assuming a mean distance of 400 pc. 
To compute the distributions, we measure the mass of the simulated central star when 1$\%$ of the envelope mass remains and compare them to the IMF of ~\citet{kroupa_variation_2001} in the range [0.01-10]~M$_{\odot}$, since the central star masses of the observed disks are not available in the two surveys reported here.
We assign a weight to each simulation accordingly, assuming a self-similar distribution of star masses across evolutionary stages.

As in Perseus (cf. Subsection \ref{subs:evtrend}), the flux emitted by the disks in Orion at millimetre wavelengths decreases as the disks evolve. 
The simulated disks emit a similar amount of flux to the Orion Class 0 disks, with a median 1.5 times lower than for the observed disks. The flux of the Stage I simulated disks overestimates the observations with a median four times larger than the observed one. Once more, we find that the simulated disks do not follow the rate of evolution suggested by the observations, since the median and distribution of simulated disks barely change between Stage 0 and Stage I. 
We also report the results for Class II disks, even if this comparison is limited by the caveats discussed in Subsection~\ref{subs:evtrend}. The median flux of the simulated Stage II disks is twenty times larger than the observed one. However, Orion A is believed to be a 1-3~Myr old cloud, while the mean age of our simulated Stage II disks is 0.5-1~Myr. The agreement gets better when considering the lower limit of the error bars.

Figure~\ref{fig:Orion_radius} shows the distribution of dust disk radius for simulated and observed disks. The dust disk radii from observations are computed in \citet{Tobin2020} as the radius at which 95$\%$ of the continuum flux is contained. In the same way, the dust disk radii from simulations are computed as the radius containing 95$\%$ of the mm flux. \citet{vanTerwisga2022} does not report size estimations and therefore we limit the comparison to the Class 0 and Class I stages. 

Contrary to what we discussed for the mm flux, the radius of the observed disks does not change significantly between Classes, similarly to what the AGE-PRO collaboration found between Class I and II disks~\citep{vioque_alma_2025}. Our simulations show a similar evolutionary trend and they can reproduce the majority of the spread of the observations.
However, we underestimate the median dust disk size of the observed objects, especially for Class 0 disks. 

\section{VLA data}
\label{Appendix c}

\begin{figure}
	\centering
	\includegraphics[width=\columnwidth]{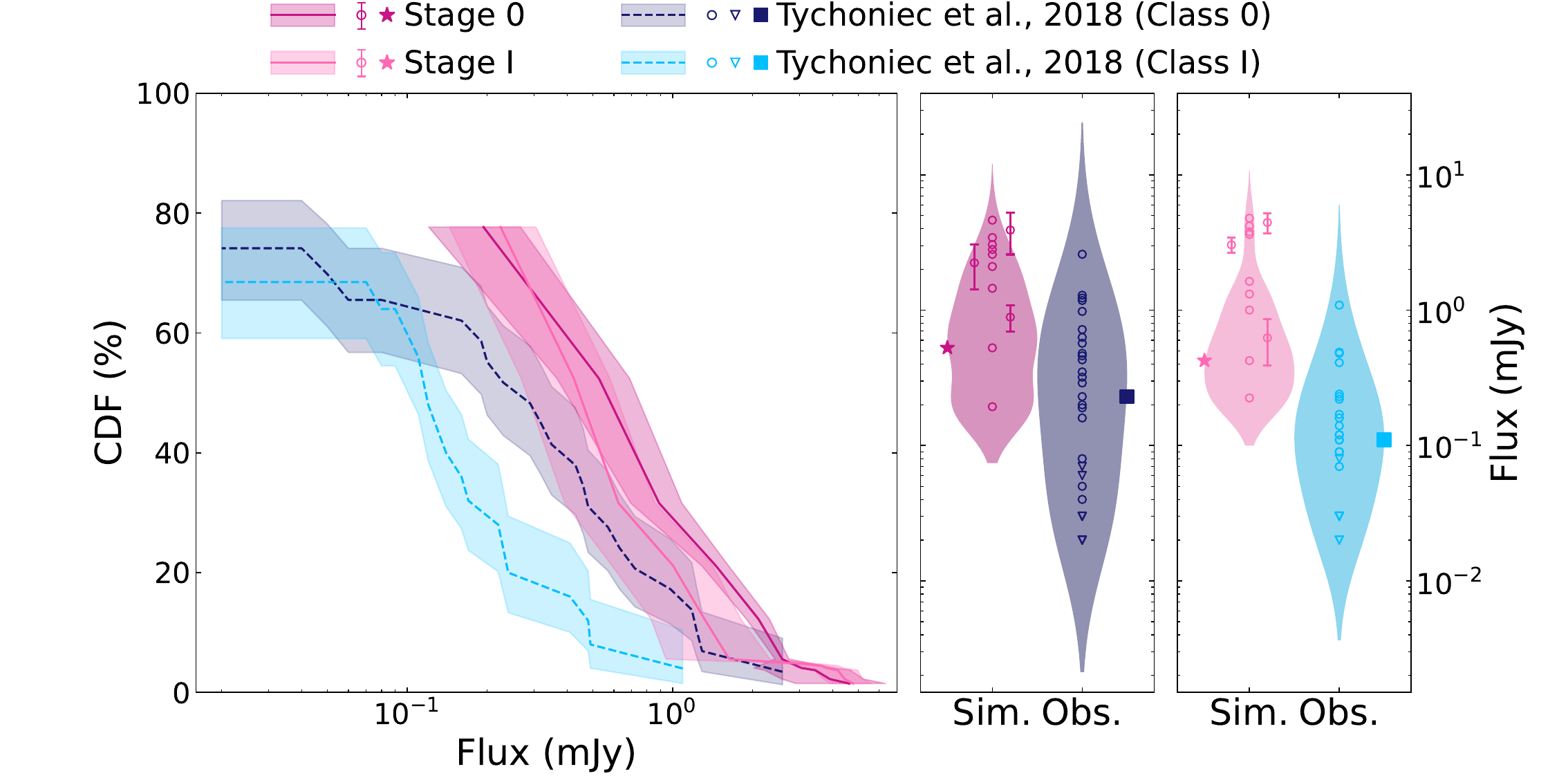}
	\caption[]{Evolutionary  trends of cm flux for disks in the Perseus star forming region vs. simulated disks. The observed fluxes have been taken from~\citet{tychoniec2018}. The colors represent the different evolutionary stages, from Class 0 to Class I, from darker to lighter. Left panel: Cumulative distribution of cm flux, with 1$\sigma$ errors shown as a shaded area for the observed disks. For the simulated disks, the shaded area shows the boundaries of our Stage 0 and Stage I definitions. Right panels: Distribution of cm flux for Class 0 (left) and Class I (right) disks. The circles show the data points, while the stars and squares show the medians, for the simulated and observed disks, respectively. The triangles show the upper limits of the observed disks as reported in~\citet{tychoniec2018}. The errorbars represent the boundaries of our Stage 0 and Stage I definitions for three representative data points (M0.40, M1.40 and M2.40).
 }
	\label{fig:Pers_VLA}
\end{figure}

Figure~\ref{fig:Pers_VLA} presents the flux emission at 9 mm for disks at two different stages of their evolution. From dark to light blue, we show the observed Class 0 and Class I disks in the Perseus molecular cloud from~\citet{tychoniec2018}, corrected for free-free emission. We exclude multiple systems with separation $<$1000~au from the sample, as reported in~\citep{Tobin2016}.

Figure~\ref{fig:Pers_VLA} also shows the results from our simulations. The cm flux has been computed using the DIANA prescription for the dust opacities and assuming a mean distance of 293 pc to Perseus. 

The flux emitted by the disks in Perseus at cm wavelengths decreases as the disks evolve, similar to what we see in Fig.~\ref{fig:Perseus_mm}, but in a less evident way. While the median of mm flux in Class I decreases more than 3 times from Class 0, here it is only halved. This difference could be due to the fact that particles may have grown to sizes that do not emit efficiently at mm wavelengths by Class I, but may still emit at cm wavelengths. 
The simulated disks emit a higher amount of flux in both Stage 0 and Stage I compared to the observed disks. However, the disagreement worsen at later stages, with the median flux from the simulated disks increasing from 2.5 times that of the observed Class 0 disks to 4 times that of the Class I disks. 
Once again, we find that our simulations are evolving more slowly than what the evolution between different Classes of observed disks suggests.


\bsp	
\label{lastpage}
\end{document}